\documentclass[fleqn,usenatbib, useAMS]{mnras}

\usepackage{newtxtext,newtxmath}

\usepackage[T1]{fontenc}
\usepackage{ae,aecompl}

\usepackage{graphicx}	
\usepackage{amsmath}	
\usepackage{amssymb}	
\graphicspath{{figures/}} 

\newcommand{\ramses}{{\sc ramses}}          
\newcommand{\mesa}{{\sc mesa}}          

\newcommand{\Fig}[1]{Fig.\ \ref{fig:#1}}    

\title[The luminosity spread in young clusters]{Explaining the luminosity spread in young clusters: \\ proto and pre-main sequence stellar evolution in a molecular cloud environment}

\author[Jensen \& Haugb{\o}lle]{Sigurd S.~Jensen\thanks{Contact e-mail: \href{mailto:sigurdsjensen@nbi.ku.dk}{sigurdsjensen@nbi.ku.dk}} \& Troels Haugb{\o}lle\thanks{Contact e-mail: \href{mailto:haugboel@nbi.ku.dk}{haugboel@nbi.ku.dk}}
\\
Centre for Star and Planet Formation, Niels Bohr Institute and Natural History Museum of Denmark, University of Copenhagen,\\
{\O}ster Voldgade 5-7, DK-1350 Copenhagen K, Denmark}

\date{Accepted 2017 October 31. Received October 30; in original form 2017 June 27}

\pubyear{2017}

\begin{document}
\label{firstpage}
\pagerange{\pageref{firstpage}--\pageref{lastpage}}
\maketitle

\begin{abstract}
Hertzsprung-Russell diagrams of star forming regions show a large luminosity spread. This is incompatible with
well-defined isochrones based on classic non-accreting protostellar evolution models. Protostars do not evolve in isolation
of their environment, but grow through accretion of gas. In addition, while an age can be defined for a star forming region,
the ages of individual stars in the region will vary. We show how the combined effect of a protostellar age
spread, a consequence of sustained star formation in the molecular cloud, and time-varying protostellar accretion for individual
protostars can explain the observed luminosity spread. We
use a global MHD simulation including a sub-scale sink particle model of a star forming region to follow the accretion process of each star.
The accretion profiles are used to compute stellar evolution models for each star, incorporating a model of how the accretion
energy is distributed to the disk, radiated away at the accretion shock, or incorporated into the outer layers of the protostar.
Using a modelled cluster age of 5 Myr we naturally reproduce the luminosity spread and find good agreement with observations
of the Collinder 69 cluster, and the Orion Nebular Cluster. It is shown how stars in binary and multiple systems can be externally
forced creating recurrent episodic accretion events. We find that in a realistic global molecular cloud model massive stars build up mass
over relatively long time-scales. This leads to an important conceptual change compared to the classic picture of non-accreting
stellar evolution segmented in to low-mass Hayashi tracks and high-mass Henyey tracks.
\end{abstract}


\section{Introduction}
Traditionally, the stellar structure of new-born stars have been calculated as a closed system, assuming initially a low surface
temperature, large radii, and a given mass. Placed on the Hayashi track the protostar first contracts at almost constant effective
temperature, and then moves towards the main sequence.
In reality, stars are born from collapsing cores in molecular clouds and grow through accretion of gas on to a
circumstellar accretion disk, where matter is subsequently transported to the star. The accretion rate starts out at a relatively
high pace and slows down while matter in first the envelope, during the protostellar phase, and then in the circumstellar disk, during
the pre-main sequence phase (PMS), is exhausted. The accretion is not smoothly decreasing but rather intermittent and unsteady in nature.
Observations of embedded protostars as well as FU~Orionis and Ex-Lupin type PMS stars show convincing evidence of dynamical
accretion both during the protostellar and PMS phase
\citep{Herbig:1966jo,Herbig:1977gf,Herbig:1989wl,1990AJ.....99..869K,2007A&A...470..211K,2009ApJS..181..321E,Scholz:2013ij,Jorgensen:2015kz,2015ApJ...800L...5S,2017ApJ...837L..29H}.
Furthermore, young clusters are observed to have a large spread in the observed luminosity in the Hertzsprung-Russell (HR) diagram.
Classical non-accreting formation models lead to well defined isochrones, which have been used for decades for age estimation of
star forming regions. However, the large luminosity spread observed in such populations has been a cause for concern.
In the classical paradigm such large spreads in luminosity imply active star formation over tens of millions of years and thus favour a
slow star formation process, which is in conflict with observed and modelled lifetimes of star formation regions \citep{2016ApJ...822...11P}.
\citet{2009ApJ...702L..27B,2012ApJ...756..118B} suggested that the apparent age spread in star forming regions could be
explained by accreting star forming models, specifically models featuring accretion bursts.

A central feature of accreting models is the energy released at the accretion shock in the boundary layers of the protostar. Potential
energy is heating the disk through vicious processes and becomes kinetic energy when matter is transported towards the star.
At the accretion shock the infall is stalled
and kinetic energy turns into thermal energy, which is either radiated away or absorbed by the protostar. The first scenario is
termed cold accretion while the second scenario where a fraction of the accretion luminosity heats the protostar is known as warm or hot accretion.
What fraction of the accretion luminosity is absorbed by the star remains an open question and probably depends on the
conditions of the shock front and the geometry of the accretion flow.
Hot and cold accretion models behave very differently and robust models must feature a reasonable approximation
for the energy dynamics at the shock front.

We present a new framework to understand the luminosity spread in young stellar clusters. As input for the stellar evolution models
we use a realistic large-scale simulation of a molecular cloud fragment (4 pc)$^3$ in size that contains a large sample of 413 protostars,
with precisely recorded mass accretion rates.  We evolve each protostar from the birth as a second Larson core to the
main sequence. To investigate the impact of the detailed accretion process we run a number of different stellar structure
evolution models for each protostar, which include different degrees of cold and hot accretion, including an accretion rate dependent model.
The HR diagrams for the synthetic clusters are compared to observations of the Collinder 69 and Orion Nebular Cluster star forming regions, and it is shown that a combination
of two effects serve to explain the observed luminosity spread. Both the non-steady accretion rate of individual protostars, and the
age spread between cluster members serves to broaden the HR diagram, compared to classical models, and bring it in agreement
with observations.

\section{Methods}
\subsection{Molecular cloud model}\label{sec:model}
The simulation is carried out with the \ramses{} code \citep{2002A&A...385..337T} modified to include random turbulence driving,
sink particles as a sub-grid model for protostars, technical improvements allowing for efficient scaling to several thousand cores,
and an improved HLLD solver that is stable in supersonic flows with high Mach numbers.
We refer the reader to \citet{2017arXiv170901078H} for an in-depth description of the simulation, and the resulting molecular cloud structure.
In that article, there is also a full account of the sink particle model, implemented physics, and code modifications compared to the public version of \ramses{}.

Our model setup is a periodic box of size (4 pc)$^3$ with a total mass of 3,000 $\mathrm{M}_{\sun}$ and an isothermal temperature of 10 K.
Assuming a molecular weight of 2.37 the average number density in the box is 795 cm$^{-3}$. The initial magnetic field strength is
7.2 $\umu$G. In our model we solve the isothermal MHD equations. Starting from a uniform gas density and magnetic field, using
a solenoidal driving on the largest scales, the RMS velocities are amplified to a sonic Mach number
$\mathcal{M}=\sigma_\textrm{v,3D} / c_s$ of approximately 10, where $\sigma_\textrm{v, 3D}$ is the
volume averaged 3D RMS velocity and $c_s$ is the isothermal sound speed. To let the turbulent flow develop and reach a statistically
steady state we first evolve the box for 20 dynamical or turn-over times, where $t_\textrm{dyn}=L_\textrm{box} / (2 \sigma_\textrm{v,3D})$,
before self-gravity and the sink particle recipe is turned on. The turbulent driving mimics the cascading external forcing from larger scales
not covered in the model. The average density, RMS velocity, and magnetic field strength in the model are
typical for parsec sized molecular clouds \citep{2009ApJ...699.1092H,2012ARA&A..50...29C}. The balance between kinetic and
gravitational energy, the virial parameter, determines the rate at which stars are formed \citep{2012ApJ...759L..27P,2017ApJ...840...48P}.
A typical definition is to use the equivalent virial number for a homogeneous sphere, $\alpha_\textrm{vir} = 5 \sigma_\textrm{v,3D}^2R / (3 GM)$.
For our model this yields $\alpha_\textrm{vir}=0.83$, where we have used $R=L_\textrm{box}/2$. This is very similar to observed values for molecular
clouds that are in the range 0.5 to 5 \citep{2013ApJ...779..185K}. The relatively low value of $\alpha_\textrm{vir}$ is probably the cause for the late-time
formation of a gravitationally bound structure in the box.

The only difference between the simulation used in this paper and in previous articles \citep{2014ApJ...797...32P,2016A&A...587A..60F}
is the imposition of a small maximum timestep at the highest level of refinement of 80 days. At every timestep we write out the sink
particle properties, resulting in 
150 GB data for 413 stars and a high-resolution account of the accretion rate for each sink particle, with up to 1 million records for the
oldest sink particle in the model. The small timestep also facilitates an accurate longtime integration of sink particle orbits in the simulation.
Finally, compared to earlier simulations, we have corrected a bug in the innermost part of the smoothed gravitational potential used for
each sink particle that could in some circumstances lead to the disruption of systems with bound sinks, when two sinks are
orbiting inside a single cell.

The simulation uses an adaptive mesh with a root grid of 256$^3$ and 6 levels of refinement reaching a maximum resolution of
$\Delta x_\textrm{min} = 4$ pc$ / 2^{14} = 50$ AU. This is enough to capture the formation process of individual stars, and the
dynamical creation of binary and multiple system through gravitational interaction. It is insufficient to resolve circumstellar disks, and
this is important to keep in mind when interpreting accretion rates. Accretion rates as reported and used in the paper are in
reality infall rates from the surrounding gas reservoir to the circumstellar disk. The real intermittency in the accretion rates
could either be higher, if disk mediated instabilities transport angular momentum in an uneven fashion from 50 AU down to the star, or
lower, if the disk acts as a buffer that smooths out sudden bursts of infall as can happen (periodically) in e.g.~binary systems.
In any case, infall rates on larger scales supplies the mass that is accreted on smaller scales, and the two must necessarily agree
in a time-averaged sense. To account for the effects of the unresolved outflows we only accrete 50\% of the mass and
momentum to the sink particle \citep[see][for an in-depth description of the sink accretion procedure]{2017arXiv170901078H}.

Refinement is based solely on density, because our main goal is to resolve the gravitational collapse. We refine
any cell at the root grid level that has a density 10 times above the average density. Then we add new levels every time the
density increases by a factor of four, in order to keep the minimum number of cells per Jeans length
\begin{equation}
L_J = \sqrt{\frac{c_s^2 \upi}{G\rho}}\frac{1}{\Delta x}
\end{equation}
constant. In our case we have $L_J > 14.4$ everywhere except at the highest level of refinement. A sink particle is created when
the gas reaches a density, at the highest level of refinement, such that $L_J < 2 $ corresponding to a number density of
$n_s > 1.7\,10^9$ cm$^{-3}$. For details of the sink particle recipe see \citet{2014ApJ...797...32P}.

We run the model for 2.55 Myr, from the time when the first star is formed. The model has a global free-fall time of
$t_\textrm{ff}=1.18$ Myr and a dynamical time $t_\textrm{dyn} = 1.08$. The model has therefore
been evolved for a bit more than two free-fall or dynamical times with star formation turned on. This allows for a well evolved stellar population
with many T-Tauri stars, and makes the model comparable in age to some of the youngest nearby star forming regions.
At the end of the run 214 stars are more than a million years old, while 328 stars have an accretion rate
that is less than $10^{-7} \mathrm{M}_{\sun}$ yr$^{-1}$, indicating that they have finished their main accretion phase. Below we make
a more detailed selection criteria and find that 182 of the stars are deeply embedded, and therefore in the Class 0 phase. The embedded
stars are mostly young and small. 40 have a mass below the brown dwarf limit of 0.076 $\mathrm{M}_{\sun}$, and 127 have a mass below
0.3 $\mathrm{M}_{\sun}$. Of the embedded stars, only 16 are more massive than 1 $\mathrm{M}_{\sun}$. The median age of the embedded
stars is 480 kyr.

\begin{figure*}
 \includegraphics[width=2.2\columnwidth]{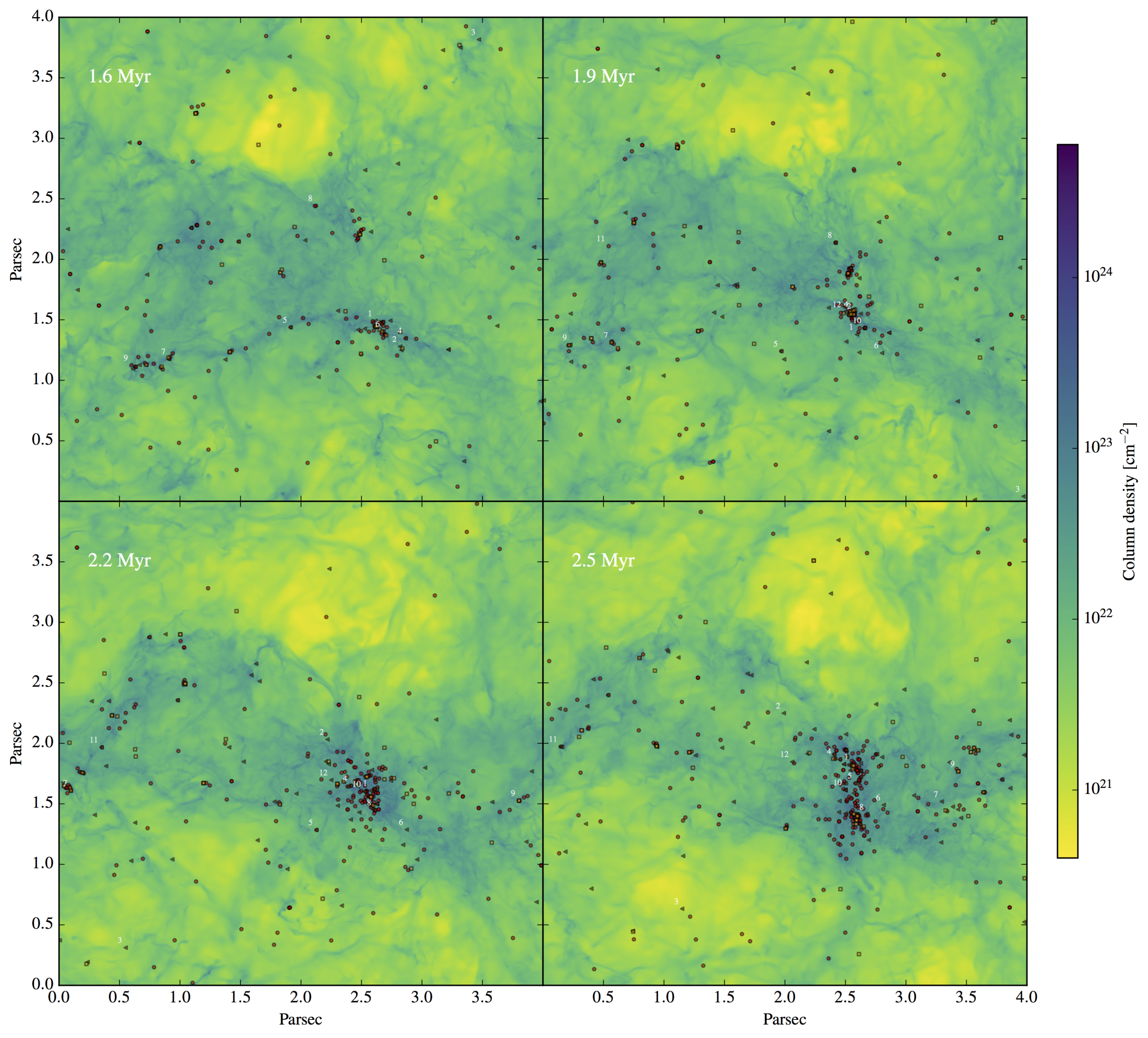}
  \caption{Column density from left to right and top to bottom at $t$ = 1.6, 1.9, 2.2 and 2.5 Myr after the formation of the first star.
  The red circles mark the positions of stars with
  masses $M < 0.5 \mathrm{M}_{\sun}$, brown triangles are stars with $0.5 \mathrm{M}_{\sun} < M < 1.5 \mathrm{M}_{\sun}$, while orange
  squares indicate stars with $M > 1.5 \mathrm{M}_{\sun}$. Numbers refer to the stars shown in \Fig{thumbs}.}
  \label{fig:cluster}
\end{figure*}

\begin{figure}
  \includegraphics[width=\columnwidth]{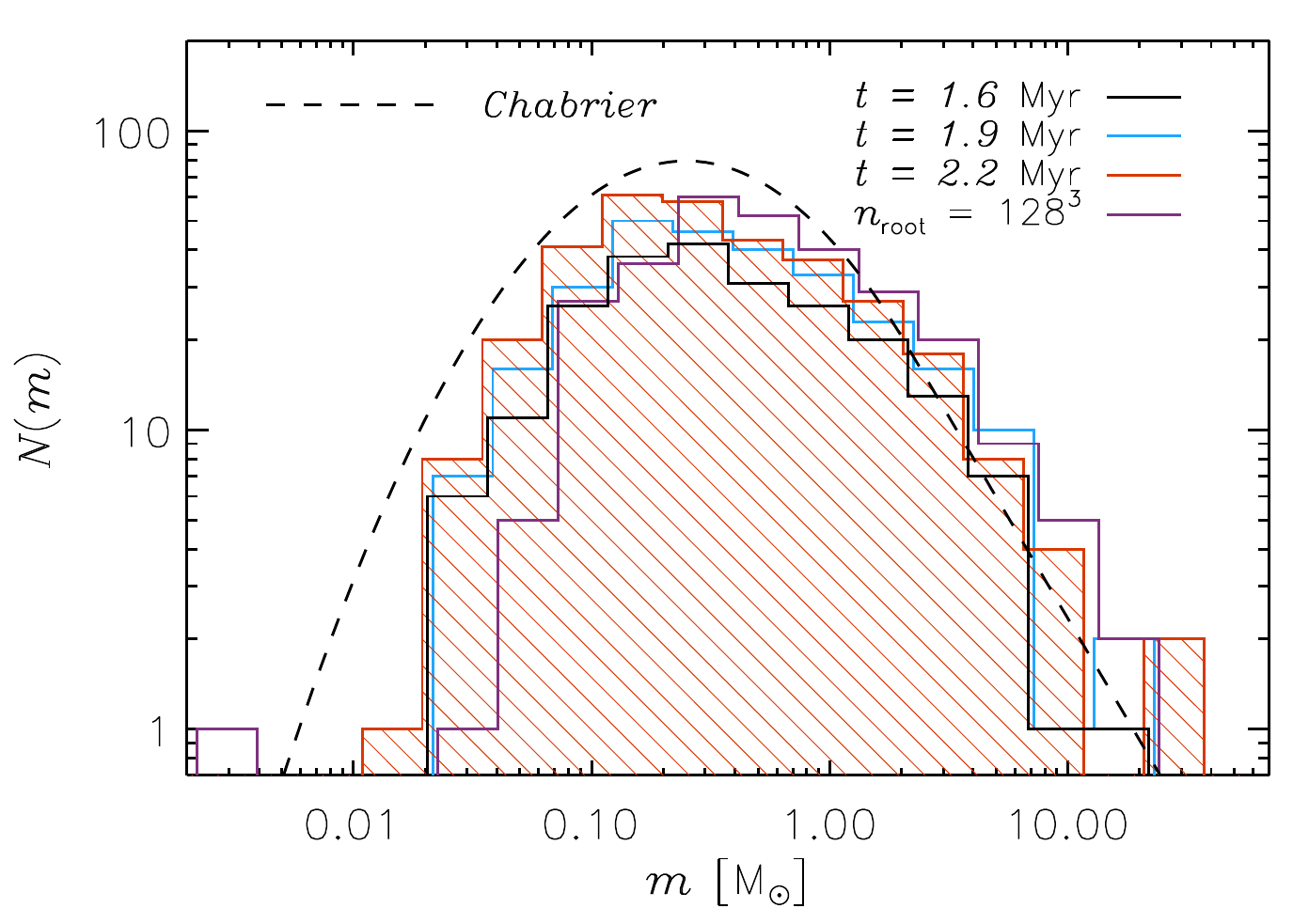}
  \caption{Initial mass function sampled at $t$ = 1.6, 1.9, and 2.2 Myr, corresponding to an SFE of 6\%, 8\%, and 10\%, and the IMF
  at an SFE of 10\% for a simulation with 8 times lower resolution. The histogram bins have been slightly offset for clarity. The dashed
  line is a Chabrier IMF normalised to the same total stellar mass as at 2.2 Myr. The vertical shifts in the histograms is due to the
  increased stellar mass as a function of time. The only significant difference between the different IMFs is the growth in the Salpeter
  range with time, which happens because the formation time of most massive stars is comparable or larger than the age of the cluster.}
  \label{fig:IMF}
\end{figure}

\begin{figure}
  \includegraphics[width=\columnwidth]{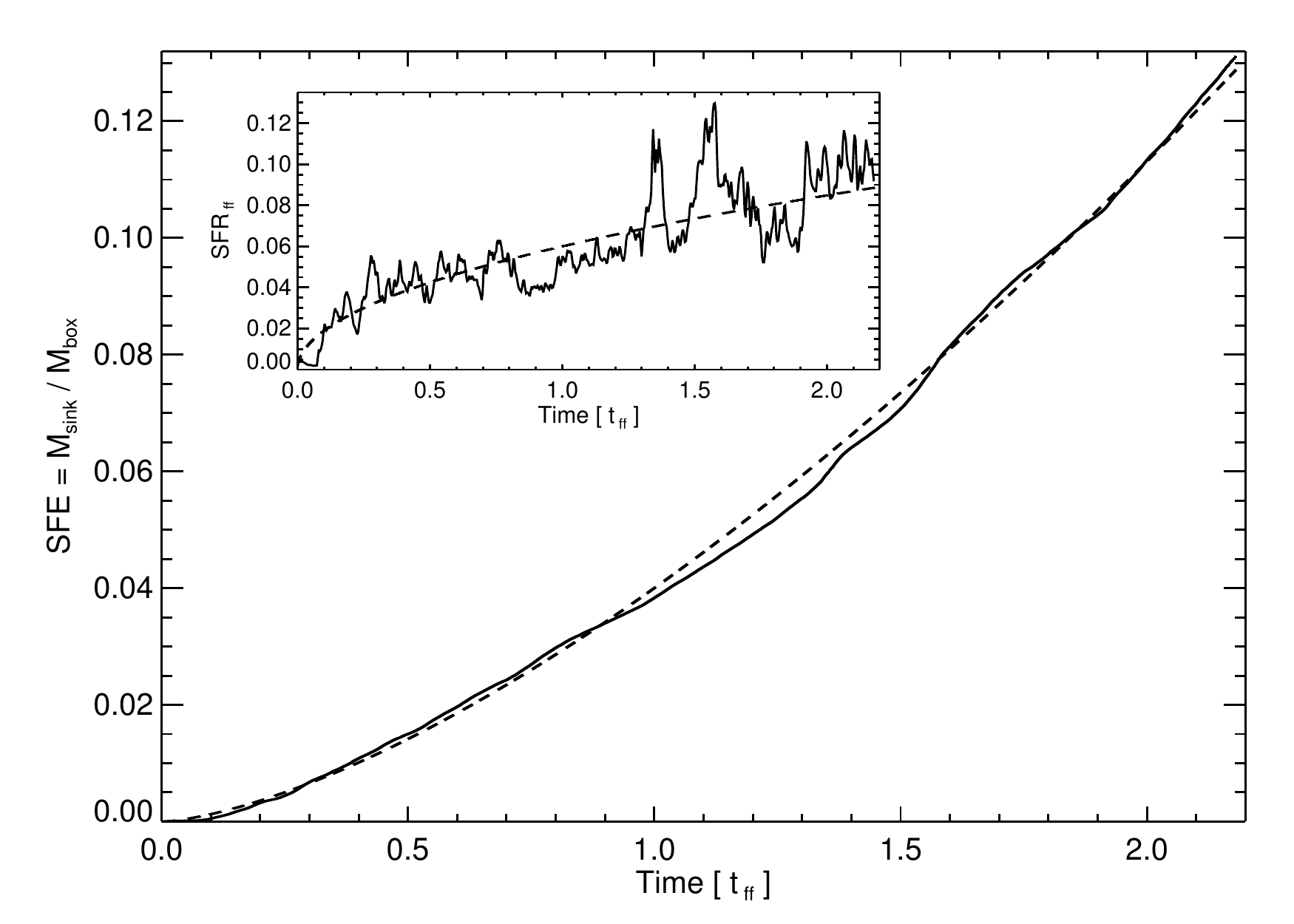}
  \caption{The evolution of the star formation efficiency and star formation rate per free-fall time (inset) as a function of time, measured in free-fall times,
  in our model. The dashed line is a power-law fit SFE $=0.04 (t / t_\textrm{ff})^{1.5}$ corresponding to SFR$_\textrm{ff} = 0.06 ( t / t_\textrm{ff})^{0.5}$.}
  \label{fig:SFE}
\end{figure}

The star formation efficiency (SFE), star formation rate per free-fall time (SFR$_\textrm{ff}$) and initial mass function (IMF) are important
indicators of the numerical fidelity and realism of the model. The star formation efficiency is the fraction of mass turned into
stars: SFE = $M_\textrm{sink} / M_\textrm{box}$, while the star formation rate per free-fall time is the SFE per free-fall time:
SFR$_\textrm{ff}= \textrm{d} \textrm{SFE} / \textrm{d} (t / t_\textrm{ff})$.
In \Fig{SFE} is shown the star formation efficiency and star formation rate per free-fall time. They are well fitted by a power-law
with SFR$_\textrm{ff} \approx 0.06 ( t / t_\textrm{ff})^{0.5}$ slowly increasing from 0.04 at 0.5 Myr (0,4 $t_\textrm{ff}$) to 0.08 at 2.4 Myr (2 $t_\textrm{ff}$).
This increase in the SFR$_\textrm{ff}$ is probably related to the formation of a single dense cluster at the end of the run
(see \Fig{cluster} for column densities at different evolution stages). The increasing density will decrease the local virial number
increasing the star formation rate \citep{2012ApJ...759L..27P,2017ApJ...840...48P}. Furthermore, at the end of the run the central
cluster is similar in size to the box and therefore the driving scale, making it impossible to disrupt it further. To continue for a longer
time period, keeping the evolution realistic, external forcing with scales beyond the 4 pc box-size would have been needed to interact
properly with the cluster, such as in \citet{2017ApJ...840...48P}. A SFR$_\textrm{ff}$ of 4\% to 8\% is realistic
for the star forming regions considered below \citep{2014prpl.conf...77P}.
In \Fig{IMF} is shown the IMF sampled at 1.6, 1.9 and 2.2 Myr, and an additional IMF extracted from an identical run, where
the only difference is an 8 times lower overall resolution. This additional run was used to validate the extent of the numerical convergence
of the model. The IMF is in good agreement with the ``Chabrier IMF'' \citep{2003PASP..115..763C}, and
the low- and intermediate-mass stellar populations are essentially time-independent
at late times, with only a vertical shift due to the growing total stellar mass and the higher end of the Salpeter slope still being developed.
Comparing the IMF at the two different resolutions we
estimate the stellar population to be complete above $\gtrsim\!0.1 \mathrm{M}_{\sun}$. To resolve the brown dwarf population would require
higher resolution to better sample the turbulence at all scales and recover the rare very high density peaks, required for the
formation of brown dwarf mass stars, but for the present purpose the cost of such a run -- $\sim$10 million cpu hours -- is prohibitive. Furthermore,
in this paper we are mainly comparing to observations of stars with masses above $\gtrsim\!0.2 \mathrm{M}_{\sun}$, and resolving the brown dwarfs population
while interesting in itself would not add additional constraints.

In this work it is important to distinguish between the cluster age, which is defined as the time since the formation of the first star
in the molecular cloud simulation, and the stellar age, which is defined from the creation of the sink particle.

Below we compare the sink particle population with clusters of an estimated age of $\sim$5 Myr. 
To extend the stellar population in our simulation to these time-scales we use the following approach. 
We consider a cluster that consists of two representations of the sink population from the molecular cloud simulation.
Each population is evolved using the stellar evolution code described in the next section. 
The first population is evolved to a cluster age of 5 Myr with 2.55 Myr of accretion histories from the molecular cloud simulation
 and 2.45 Myr of continued evolution as non-accreting stars.
 The second population is evolved to a cluster age of 2.45 Myr and is thus still within the timespan of the molecular cloud simulation 
 allowing sink particles to be accreting, something which is essential to this work.
In this way we construct an approximation of the stellar inventory of the cluster at 5 Myr
under the assumption of continuous star formation occurring in the cluster. 
Given the time-stability of the IMF this is a reasonable approach. 
The SFR$_\textrm{ff}$ is not constant throughout the simulation, but the time variation in the SFR$_\textrm{ff}$ is acceptable.
The current run has already reached an SFE of 13\%. Stitching two copies of the stellar evolutionary profiles together, with the
the older profile only containing non-accreting stars, carries the implicit assumption that star formation continues for up to 5 Myr.
A larger box-size would be required to properly explore this, but continuing running the model to 5 Myr is, apart from costly -- it would cost
$\sim$3 million cpu hours -- also unrealistic. If the SFE continues evolving according to the power-law fit, the cloud would reach an SFE of
30\%, and protostellar feedback could become important. The typical life-time of a molecular cloud is on the order
$\sim 2\,-\,5\,\,t_\textrm{dyn} = 2\,-\,5$ Myr, and external feedback and infall from larger scales would play an important role in the
further evolution of the cloud \citep{2014prpl.conf....3D,2016ApJ...822...11P}. 

In the rest of the paper we will refer to sink particles as ``stars''.

\subsection{Stellar evolution models with \mesa{}}
The evolution of individual protostars are modelled in the one-dimensional stellar evolution code
\mesa{} version 8845. We refer to the instrument papers of \mesa{} \citep{2011ApJS..192....3P,2013ApJS..208....4P,2015ApJS..220...15P}
for a complete description of the stellar evolution code.
The code has been augmented with modules for accretion of mass with a variable entropy and diagnostics to calculate the accretion luminosity.
The implementation of the accretion luminosity is described in detail below. We do not include rotation in the models.

Each protostar is evolved from the same initial setup and with the same physics implementation.
The physical parameters were chosen to be suitable for stars of low- and intermediate masses.

Convection is implemented using the mixing length theory adopted from \citet{1968pss..book.....C} with
$\alpha_{MLT} = 1.82$.
We adopted overshooting parameters for low- and intermediate-mass stars from \citet{2016ApJ...823..102C} above and below convective zones.
Convective stability is determined based on the Schwarzschild criterion. The Ledoux criterion for convection was tested and no notable difference between the
two criteria were found for the protostellar models.
Similarly a number of alternative implementations of the mixing length theory included in \mesa{} have been tested and found to produce similar results.

The reaction networks for $^{2}\mathrm{H}$, $^{7}\mathrm{Li}$, $^{7}\mathrm{Be}$ and $^{8}\mathrm{B}$ were enabled
in addition to the standard nuclear reaction network of \mesa{}.

\mesa{} allows for different opacity tables in different temperature regimes.
In the low temperature regime ($T < 4,000$ K) we use opacity tables from \citet{2008ApJS..174..504F} and for higher temperatures
the tables of \citet{1998SSRv...85..161G} is used.
We use the standard \mesa{} equation of state (EOS), which is based on the OPAL EOS tables of \citet{2002ApJ...576.1064R} and
extended to lower temperatures with the tables of \citet{1995ApJS...99..713S}.

\subsubsection{Initial conditions}
The initial setup for the \mesa{} models is a relaxed second Larson core of uniform composition.
Recent studies of the collapse of prestellar cores with a broad spectrum of different initial conditions
found second Larson cores with $M \sim 0.0029 \mathrm{M}_{\sun}$ and $R \sim 0.82 \mathrm{R}_{\sun}$ to be universal for
low- and intermediate-mass stars \citep{2017A&A...598A.116V}
and we have chosen to use cores with this mass for our simulations.
We have not fixed the initial radius and due to the differences between the opacity tables we have utilised in \mesa{} and those of \citet{2017A&A...598A.116V} we get an initial radius of $R = 0.64 \mathrm{R}_{\sun}$.

The composition of the core is uniform and the mass fractions for the essential elements of the protostar are
based on the values for the local interstellar medium \citep{2001RvMP...73.1031F, 2000IAUS..198..525T,2010MNRAS.406.1108P}:
$X=0.70$, $Y = 0.28$, $^{2}\mathrm{H} = 2\times10^{-5}$ and $^{3}\mathrm{He} = 2.98\times10^{-5}$.
For the heavier elements we use the mass fractions of \citet{1998SSRv...85..161G}.
We accrete matter with the same composition as the initial core.

\subsubsection{Modeling energy transfer at the accretion shock}
The accretion shock near the surface of the protostar is a crucial part of accreting star formation models,
but our understanding of the energy transfer at the shock remains limited.

We have adopted the terminology of \citet{2009ApJ...702L..27B} for the energy transfer at the accretion shock.
The accretion luminosity is divided into two terms, one for the energy radiated away from the protostar
and one for the energy that is injected into the protostar.
\begin{equation}
L_{\mathrm{acc}}^{\mathrm{out}} = \epsilon (1-\alpha) \frac{G M \dot{M}}{R} \quad L_{\mathrm{acc}}^{\mathrm{in}} = \epsilon \alpha \frac{G M \dot{M}}{R}
\end{equation}
Where $M$ and $R$ are the protostellar mass and radius respectively, $\alpha$ is the thermal efficiency parameter,
$G$ is the gravitational constant and $\dot{M}$ is the accretion rate.
The factor $\epsilon$ depends on the details of the accretion process, with $\epsilon \leqslant 1$ for gravitationally bound material and $\epsilon \leqslant 0.5$ for boundary layer accretion from a thin disk.
We set $\epsilon = 0.5$ as we assume thin disk accretion as previously done by \citet{2009ApJ...702L..27B}.
This requires that half of the accretion luminosity is stored in rotational energy or lost during accretion in the disk.
This would be the case if e.g.~the distance between the inner gap of the accretion disk and the stellar surface is one stellar
radius and that it coincides with the co-rotation radius \citep{2016ARA&A..54..135H, hartmann_2008}.
The size of the gap can be measured indirectly by
assuming co-rotation between the foot points of magnetospheric accretion channels originating close to the
inner edge and the PMS star, and support such a size range \citep{2007IAUS..243..135C,2010ApJ...718..774E}.
In the earliest stages of star formation before an accretion disk has formed and has become the dominant means
of transferring mass from the envelope to the protostar this approximation
is not valid, which causes the accretion luminosity we calculate in that case to be a lower limit. This is also the case if
the foot points of the accretion channels or the co-rotation radius are further away than two stellar radii.

Spherical accretion without rotation was studied extensively by \citet{1980ApJ...241..637S, 1980ApJ...242..226S, 1981ApJ...248..727S}
and they found that for an optically thick accretion shock the maximum thermal efficiency is 75\%
\footnote{In the case of $\epsilon = 1.0$ then $\alpha = 0.75$}.
Observations of T-Tauri stars have indicated that a large fraction of the gas accreted in this evolutionary stage happens
through an accretion disk. \citet{1997ApJ...475..770H} argue that accretion through a disk must be cold or have a very low
thermal efficiency. This is due to the optically thick nature of the protostar compared to the surrounding regions,
into which the accretion luminosity can radiate freely when accretion occurs on a limited region of the protostellar surface
through e.g.~magnetospheric accretion channels.

\citet{2016A&A...588A..85G} recently compared two-dimensional simulations of accreting young stars with previous
one-dimensional models of accreting protostars.
They found a reasonable agreement between the different models while there was a tendency for the one-dimensional models
to exaggerate the effects of a high thermal efficiency relative to the two-dimensional models.

Below, we present results for a number of constant thermal efficiencies as well as an dynamic $\alpha$-model in which the
thermal efficiency is a function of the instantaneous accretion rate.
The model is motivated by the assumption that the accretion flow must cover the majority of the protostellar surface during
high accretion rates leading to a high thermal efficiency, while accretion at lower rates is expected to be more localised occurring
for example as magnetospheric accretion through a magnetised loop in which case material will be aggregated to the stellar
surface  at a low thermal efficiency with the same entropy as surface material.

As a result of the work presented by \citet{2016A&A...588A..85G} we have limited $\alpha \eqslantless 0.5$.
Together with the choice of $\epsilon = 0.5$ this means that at most 25\% of the accretion energy is absorbed by the protostar.
This is a conservative upper limit for the thermal efficiencies and the effect of hot accretion can be considered moderate in this case.
A higher thermal efficiency leads to a larger protostar as the continuous addition of energy alters the thermal structure of the protostar.
This results in a longer contraction phase towards the main sequence once accretion is stopped and delays the onset of nuclear processes
such as deuterium and lithium burning.

The dynamic $\alpha(\dot{M})$-model is constructed as a step function with a smooth transition. It is given by the following expression
\begin{equation}
\alpha(\dot{M}) = \frac{\alpha_{L} \exp (\frac{\dot{M}_{m}}{\Delta}) + \alpha_{H} \exp (\frac{\dot{M}}{\Delta})}{\exp (\frac{\dot{M}_{m}}{\Delta}) + \exp ( \frac{\dot{M}}{\Delta})}\,,
\end{equation}
where $\alpha_{L} = 0.005$, $\alpha_{H} = 0.5$ are the lower and upper bounds of $\alpha$ and
$\dot{M}_{m}=3.5\times10^{-5} \mathrm{M}_{\sun} \mathrm{yr}^{-1}$ and $\Delta=(2.0/3.0)\times10^{-5} \mathrm{M}_{\sun} \mathrm{yr}^{-1}$ are the midpoint
and the width of the crossover between $\alpha_{L}$ and $\alpha_{H}$.

For the case of hot accretion we have implemented a module which calculates $L_{\mathrm{acc}}^{\mathrm{in}}$ and distributes the energy across the
outer layers of the protostar.
Each cell in the deposition region receives the same specific energy addition such that the energy is distributed evenly with respect to the stellar mass.
For numerical stability, we set an upper limit on the specific energy addition in each cell, such that a higher
$L_{\mathrm{acc}}^{\mathrm{in}}$ results in distribution across a larger number of cells and thus a deeper deposition of energy.
We have tested the robustness of this implementation by varying the energy limit for each cell so that the energy is distributed
across larger and smaller regions and found almost no variation in the evolution of the protostars, given that convection in the surface
layers in most cases readily redistribute the added heat.

 \begin{figure*}
  \includegraphics[width=\textwidth]{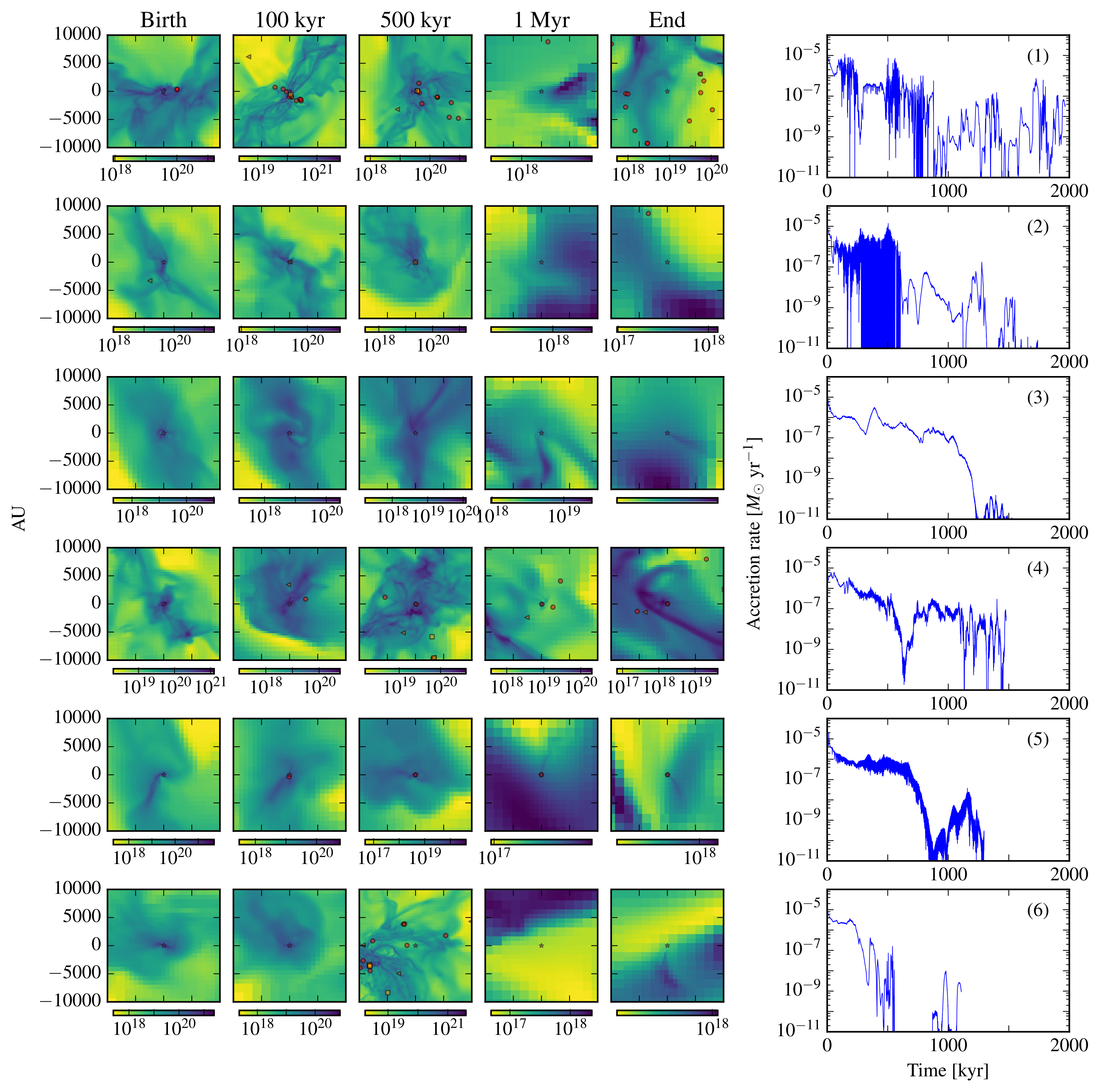}
  \caption{Thumbnails for 12 different stars (one per row, continued on next page) with a mass between 0.7 and 0.8 $\mathrm{M}_{\sun}$ at
  a cluster age of 2.3 Myr. The numbers in parenthesis indicated
  in the rightmost plot correspond to the numbers in Figs.~\ref{fig:cluster}, \ref{fig:sim-mass1}, and \ref{fig:sim-mass2}. Each column
  density plot is based on a (10,000 AU)$^3$ cut-out from the model. From left to right the panels show the gas distribution at the birth
  of the star, after 100 kyr, 500 kyr, 1 Myr, and at a cluster age of 2.3 Myr. The magenta star at the center of each
  panel indicates the star position, red circles are stars with masses $M < 0.5 \mathrm{M}_{\sun}$, brown triangles are stars with
  $0.5 \mathrm{M}_{\sun} < M < 1.5 \mathrm{M}_{\sun}$, while orange squares indicate stars with $M > 1.5 \mathrm{M}_{\sun}$. The right-most panel shows
  the accretion rate. The panels illustrate the effect of different environments and the variability of accretion
  histories in a single mass bracket. Notice how the presence of binary and multiple systems in the model naturally induces
  rapidly varying accretion rates for some of the stars.}
  \label{fig:thumbs}
\end{figure*}

\begin{figure*}
  \includegraphics[width=\textwidth]{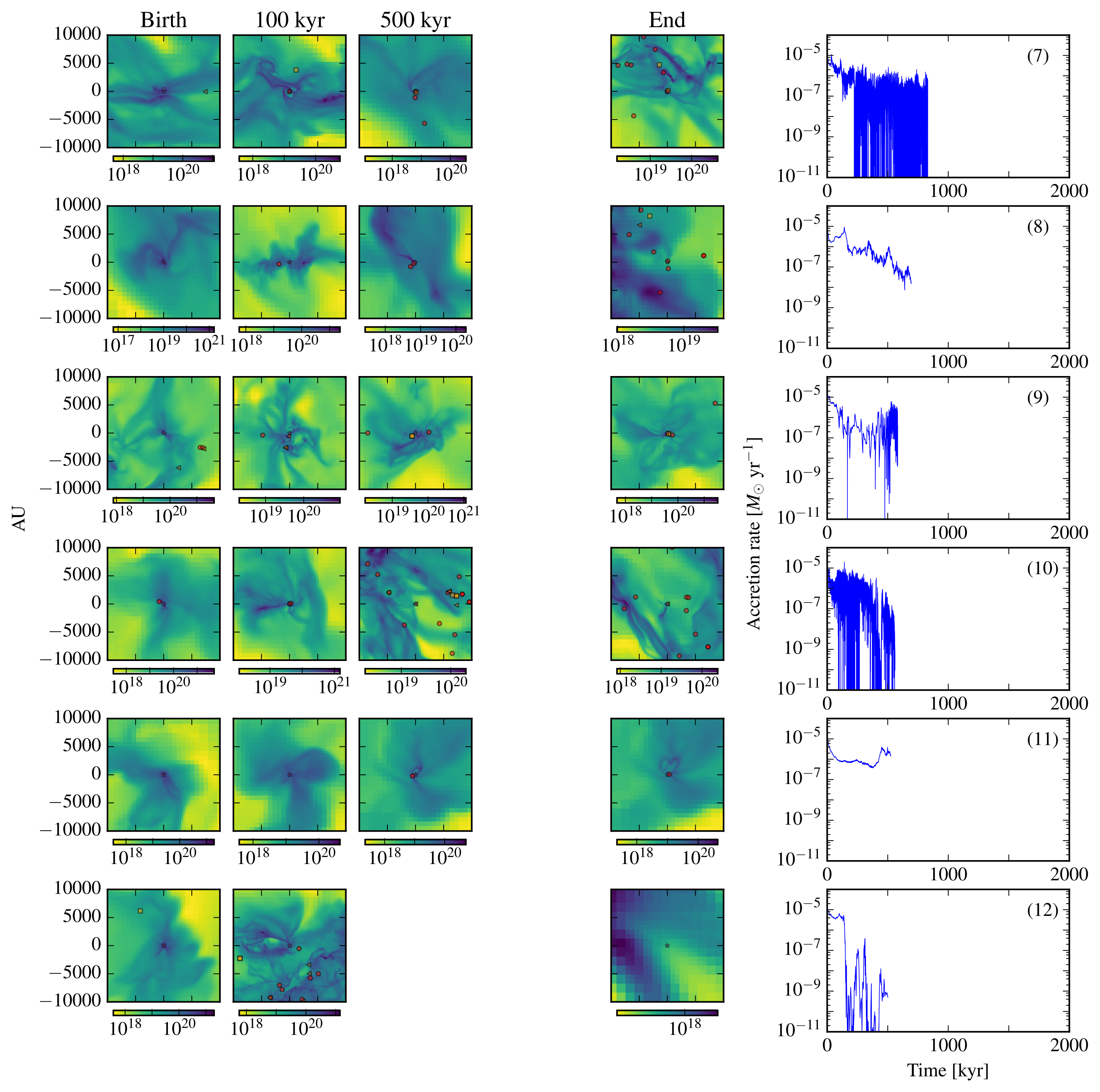}
  \contcaption{from previous page}
  \label{fig:thumbs-continued}
\end{figure*}

\section{Results}
In this section we present the stellar evolution models, their tracks in the HR diagram, and compare models with different thermal efficiencies.

\subsection{Protostellar models and thermal efficiency parameters}\label{sec:HRtracks_alpha}
\begin{figure}
  \includegraphics[width=\columnwidth]{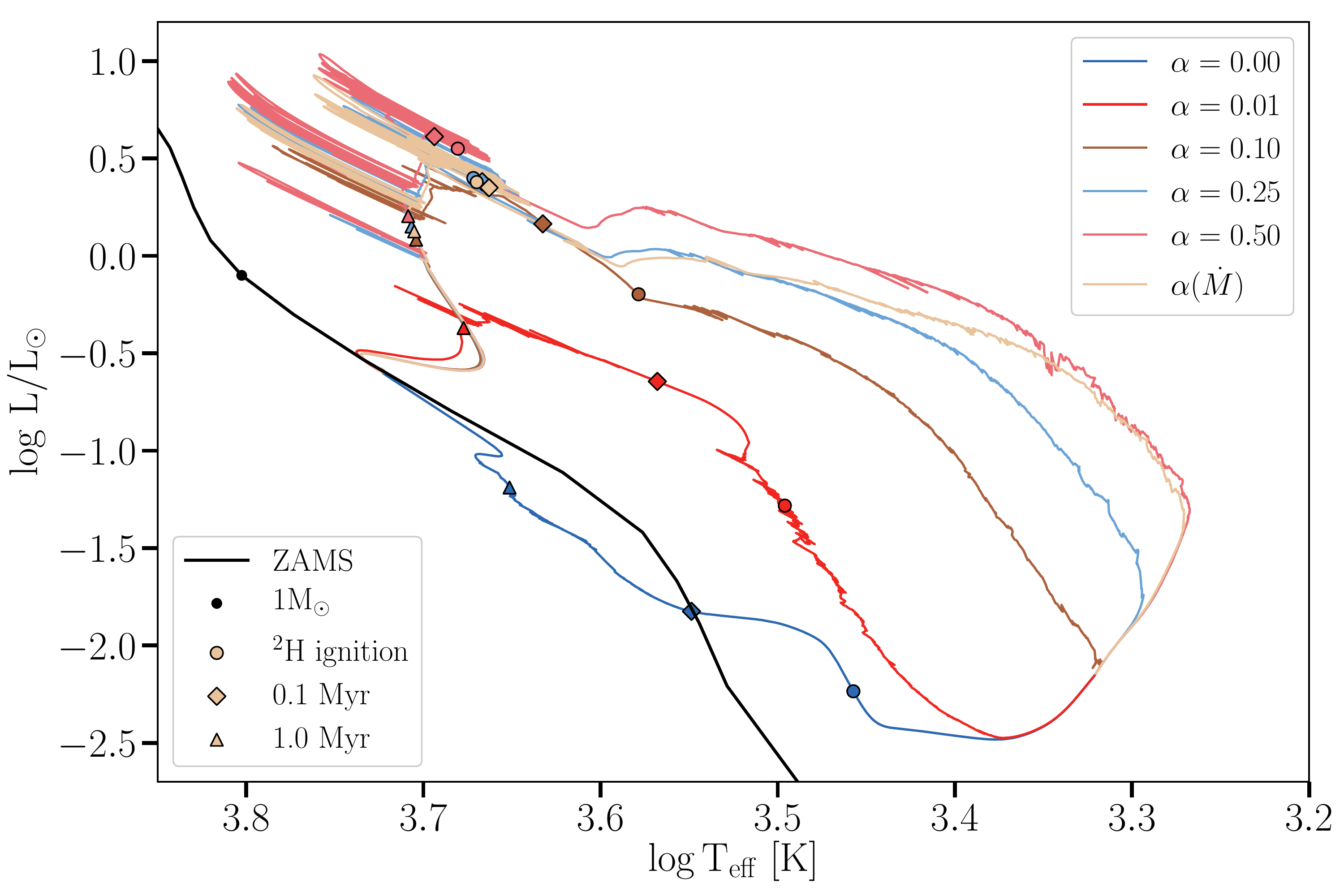}
  \caption{Models with varying thermal efficiencies for star $\#1$ in \Fig{thumbs} with $M = 0.78 \mathrm{M}_{\sun}$.
  The outward component of the accretion luminosity $L_{\mathrm{acc}}^{\mathrm{out}}$ is not included. The environment and accretion
  profile for this star is shown in the top row of \Fig{thumbs}.}
  \label{fig:alpha-comparison}
\end{figure}

In \Fig{alpha-comparison} is shown the evolutionary tracks for a protostar with a final mass $M = 0.78 \mathrm{M}_{\sun}$ for six different thermal efficiencies.
The accretion rate of the protostar is shown in the top row of \Fig{thumbs}.
The figure includes indicators for the ignition of deuterium as well as the position of the protostar at 0.1 Myr and 1 Myr.
We note that many of the stars remain deeply embedded for the first $\sim$500 kyr and therefore the initial part of the
tracks presented here should not be compared directly with the observations. Below we use a simple criteria to
decide when a star is embedded based on a comparison between the stellar and envelope mass based on the results
of \citet{2016A&A...587A..59F}. For a more
refined estimate, a synthetic observation would have
to be performed on the simulation, propagating the luminosity calculated from the stellar model out through the cloud
environment \citep{2016A&A...587A..59F,2016A&A...587A..60F,kuffmeier}.
If we disregard anisotropic shadowing and scattering of the light in the disk and envelope the luminosity is (isotropically) preserved,
while the spectral energy distribution and effective temperature is redistributed to longer wavelengths.
We have not added the accretion luminosity to the bolometric luminosity in \Fig{alpha-comparison} to highlight the differences in
the protostellar evolution. Including $L_{\mathrm{acc}}^{\mathrm{out}}$ reduces the relative differences between the tracks as the accretion
luminosity dominates the protostellar luminosity during periods of high accretion rates.

It is evident that the initial evolution of the protostar is highly sensitive to the thermal efficiency $\alpha$.
Models with a non-zero thermal efficiency above $0.1$ undergo a rapid initial expansion phase, where the hydrostatic structure
of the star is readjusted due to the large deposition of accretion energy during the initial phase where accretion rates are high.
This effect makes the evolution in hot models insensitive to the exact size and mass choice of the initial second Larson core
as the entropy in the star is largely dominated by the accreted matter, and the hydrostatic equilibrium becomes independent
of the initial characteristics, only depending on the thermal efficiency and the initial accretion rate.
For models with $\alpha < 0.1$ the addition of matter leads to a contraction of
the protostar and they quickly reach central temperatures high enough for the ignition of deuterium and subsequently hydrogen.
The readjustment leads to a strong dependence on the initial conditions for the cold models as discussed by \citet{2011ApJ...738..140H}.
Cold models with a larger initial core will not contract enough to reach the Zero-Age Main Sequence (ZAMS) before accretion
ceases and thus ignition of hydrogen is delayed compared to the cold models presented here, which ignite hydrogen after roughly $100$ kyr while accretion is still ongoing.
The early ignition of hydrogen means that the models reach the ZAMS early on and subsequently cross it as the
continued accretion causes the models to have smaller radii than ZAMS stars of equivalent mass.
Thus the tracks of the cold models cross below the ZAMS as they continue to accrete mass and climb toward higher luminosities and
higher effective temperatures.
Once the mass exceeds $M \sim 1.4 \mathrm{M}_{\sun}$ or accretion stops the tracks rejoin the ZAMS again, where they will settle at the
locations corresponding to their final mass. An example of such an evolutionary track is seen in \Fig{alpha-comparison}.
Our model features smaller initial cores than previous studies \citep{2012ApJ...756..118B, 2017A&A...597A..19B, 2017A&A...599A..49K} of
accreting models and as a result the tracks for the cold models are different than previous studies as the protostars remain more compact.

The cold accretion models presented here follow almost identical tracks in the HR diagram independent of the accretion
rate and final mass when the accretion luminosity is not included. 
This behaviour is similar to the models of \citet{2011ApJ...738..140H} albeit the tracks are different due to different initial conditions.
Including the accretion luminosity leads to a spread in luminosities due to differences in accretion rates, but the effect is
diminished as accretion rates drop below $10^{-6} \mathrm{M}_{\sun} \mathrm{yr}^{-1}$.
We are not convinced that the evolution of these purely cold models with $\alpha = 0.0$ throughout the entire accretion phase are representative of real protostars. Physically, back-of-the-envelope calculations 
suggest that for accretion rates $\dot{M} > 10^{-5} \mathrm{M}_{\sun} \mathrm{yr}^{-1}$ some fraction of the accretion luminosity will heat the protostar (see Appendix B in \citet{2012ApJ...756..118B}). Similarly, a recent numerical study of the accretion shock onto accreting protoplanets indicate that cold accretion is unlikely for canonical values for the planetary accretion process \citep{2017ApJ...836..221M}.
Due to the compact structure, the purely cold models lack a contraction phase at the end of accretion. This leads to a bimodal feature as the cold accreting PMS stars are either in a high luminosity state -- due to
the large accretion luminosities -- or in a low luminosity state lying at or close to the ZAMS.
Such a bimodal feature does not fit observations of young clusters, which show signs of a longer contraction phase towards the ZAMS. 
We stress that we do not exclude that possibility that $\alpha = 0.0$ for large parts of the accretion process where accretion rates are below $\sim 10^{-5} \mathrm{M}_{\sun} \mathrm{yr}^{-1}$ and it is possible that a small number of very low-mass objects 
never reach accretion rates of this order and evolve through cold accretion.

The choice of thermal efficiency has another profound effect on the models presented here.
Purely cold models are fully convective until the ignition of hydrogen at which point deuterium burning continues in a shell
and the region between the burning zones is radiative.
Outside the deuterium burning shell the protostars remain convective until they settle on the ZAMS at the end of
accretion\footnote{Stars which reach central temperatures sufficient for CNO to be the predominant fusion process
become radiative in the envelope as classical theory prescribe.}.

Models with non-zero $\alpha$ exhibit only partial convection during the protostellar accretion phase. In a very small region in the center
there exist a radiatively dominated core, while high accretion rates can increase temperatures in the outermost layers of the star. This drives
a small temporary radiative zone in the middle of the star and a dynamical instability, as discussed below.
That convection is not stable in the outer layers is not surprising given the addition of entropy during hot accretion which can
alter the temperature gradient and thus inhibit the convective motion.
This effect is reproduced in two dimensional models of accreting protostars \citep{2016A&A...588A..85G}.
Despite the appearance of temporary radiative zones in the outer layers, accreted deuterium and lithium will reach the burning region and once
the star reaches the PMS these regions disappear, except when the star is in a high accretion state, as seen in \Fig{trho}.
The radiative nature of the innermost core during hot accretion is more surprising.
The same behaviour was found by \citet{1980ApJ...241..637S} and it leads to a temperature inversion at the boundary between
the radiative core and the convective exterior region, which subsequently results in a shell ignition of deuterium.
\citet{1980ApJ...241..637S} explained the occurrence of a temperature inversion between the convective region and the radiative core
by the convectively stable structure of their initial hydrostatic core. While in our models the initial core is convective at the beginning of the simulation,
the addition of accretion energy in the hot accretion models changes the temperature gradient drastically from the onset of accretion, and in
particular in the early phase. This effect causes convection to halt near the boundary of the initial core and the region remains convectively stable
until the end of the main accretion phase.
Once accretion ceases the protostar quickly reaches the same convective layout as main sequence stars of equal mass.
\Fig{trho} show star $\#1$ from \Fig{thumbs} during the main accretion phase after the onset of deuterium fusion, but just after a period with
high accretion rates. The temperature inversion is clearly visible in the high density region, where the star is radiative.

\begin{figure}
  \includegraphics[width=\columnwidth]{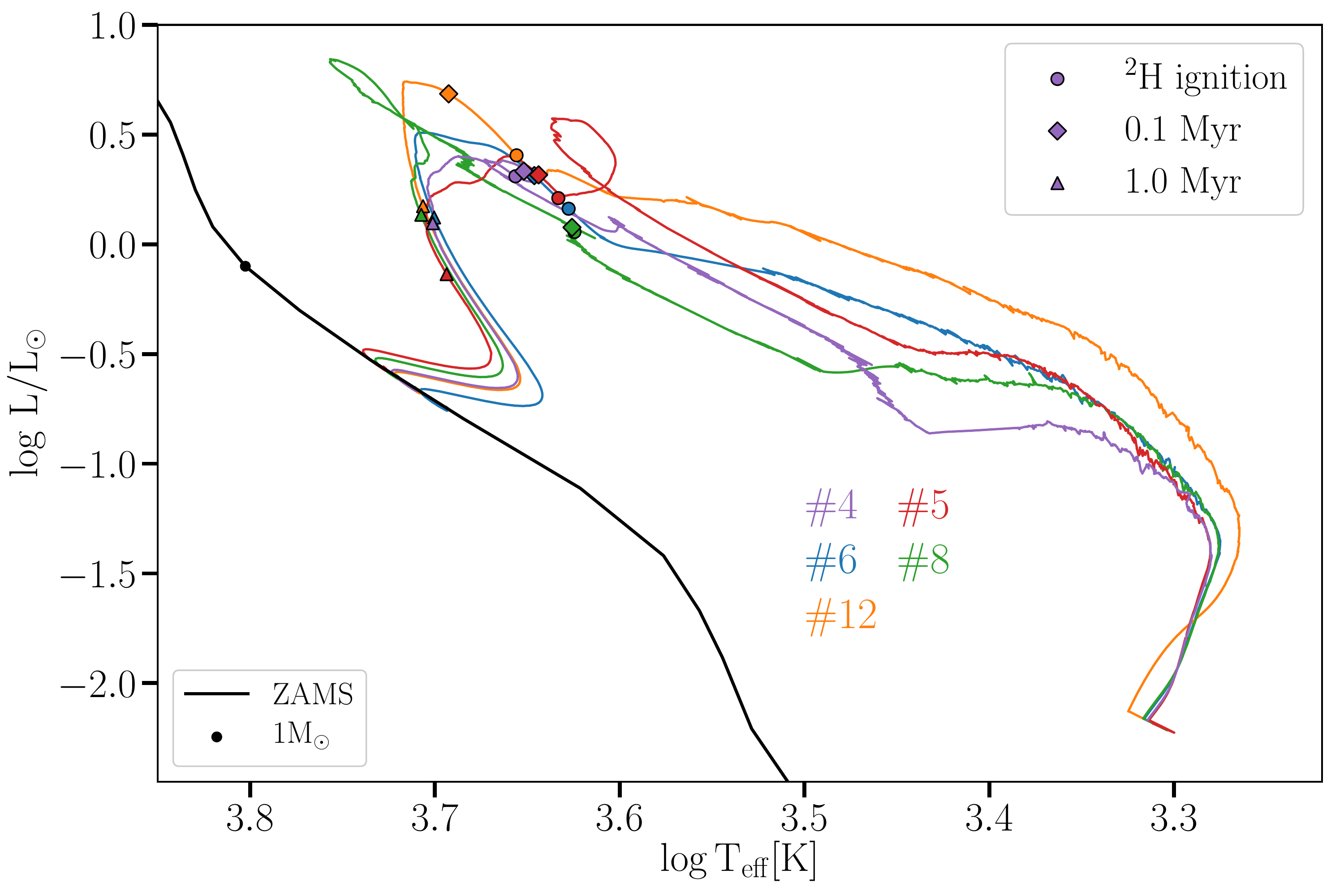}
  \caption{Evolutionary tracks for 5 stars with masses in the interval $M \in [0.7,0.8] \mathrm{M}_{\sun}$ with the dynamical thermal efficiency $\alpha(\dot{M})$.
  The stars
  in this figure does not feature significant accretion burst in the PMS phase, opposed to the stars in \Fig{sim-mass2}.
  The outward component of the accretion luminosity $L_{\mathrm{acc}}^{\mathrm{out}}$ is not included. The numbers refer to
  \Fig{thumbs}.}
  \label{fig:sim-mass1}
\end{figure}

\begin{figure}
  \includegraphics[width=\columnwidth]{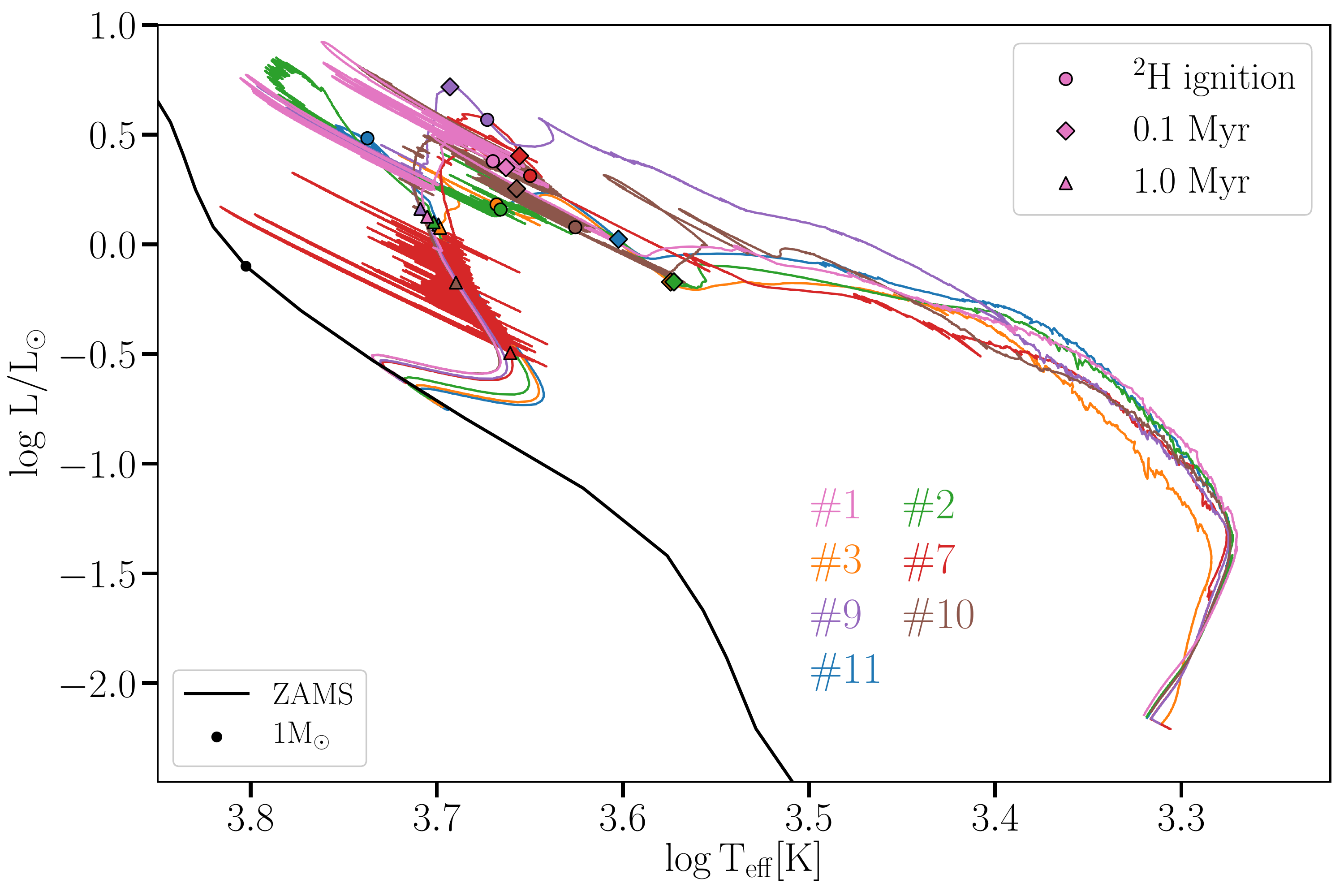}
  \caption{Evolutionary tracks for 7 stars with masses in the interval $M \in [0.7,0.8] \mathrm{M}_{\sun}$ with the dynamical thermal efficiency $\alpha(\dot{M})$.
  The stars in this figure have significant accretion bursts that cause instabilities to occur.
  The outward component of the accretion luminosity $L_{\mathrm{acc}}^{\mathrm{out}}$ is not included. The numbers refer to
  \Fig{thumbs}.}
  \label{fig:sim-mass2}
\end{figure}

In Figs.~\ref{fig:sim-mass1} and \ref{fig:sim-mass2} we present the evolution of the 12 protostars from \Fig{thumbs} in the HR diagram to
illustrate how the evolution of protostars in the same mass bracket is affected by differences in the time-dependent accretion.
The HR diagram does not include the accretion luminosity $L_{\mathrm{acc}}^{\mathrm{out}}$ to highlight the protostellar variations.
The luminosity of the protostars vary by almost 1 dex during the initial 1 Myr of the stellar life but converge to similar tracks as
they reach the contraction phase and descend on the Hayashi track.

\begin{figure*}
  \resizebox{\hsize}{!}{\includegraphics{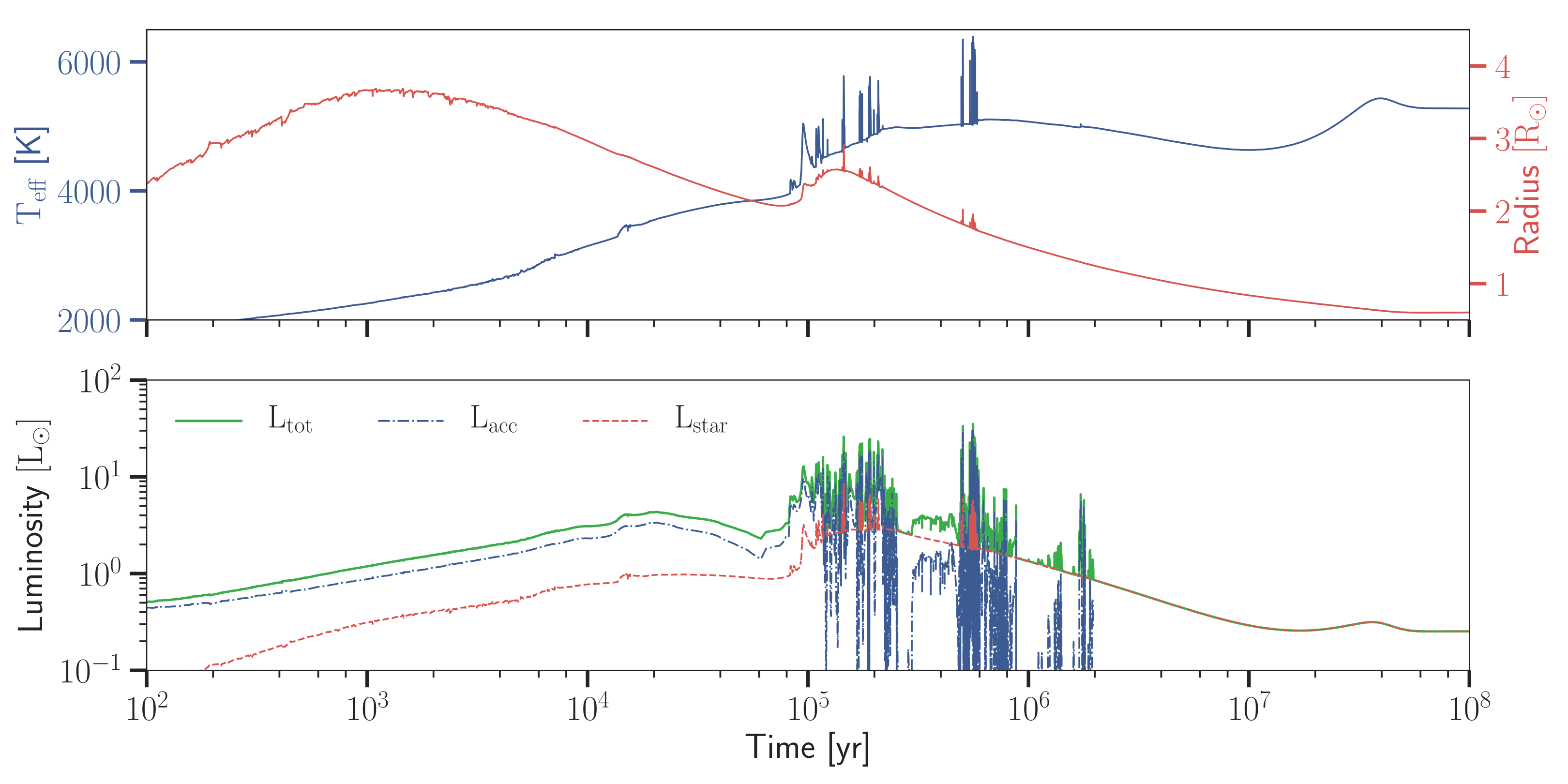}}
  \caption{The time evolution of stellar variables $T_{\mathrm{eff}}$, $R$ and the luminosity including the individual components $L_{*}$ and $L_{\mathrm{acc}}$ for star $\#1$ in \Fig{thumbs} with the dynamical thermal efficiency $\alpha(\dot{M})$.
  }
  \label{fig:sink_0025}
\end{figure*}

\begin{figure}
  \includegraphics[width=\columnwidth]{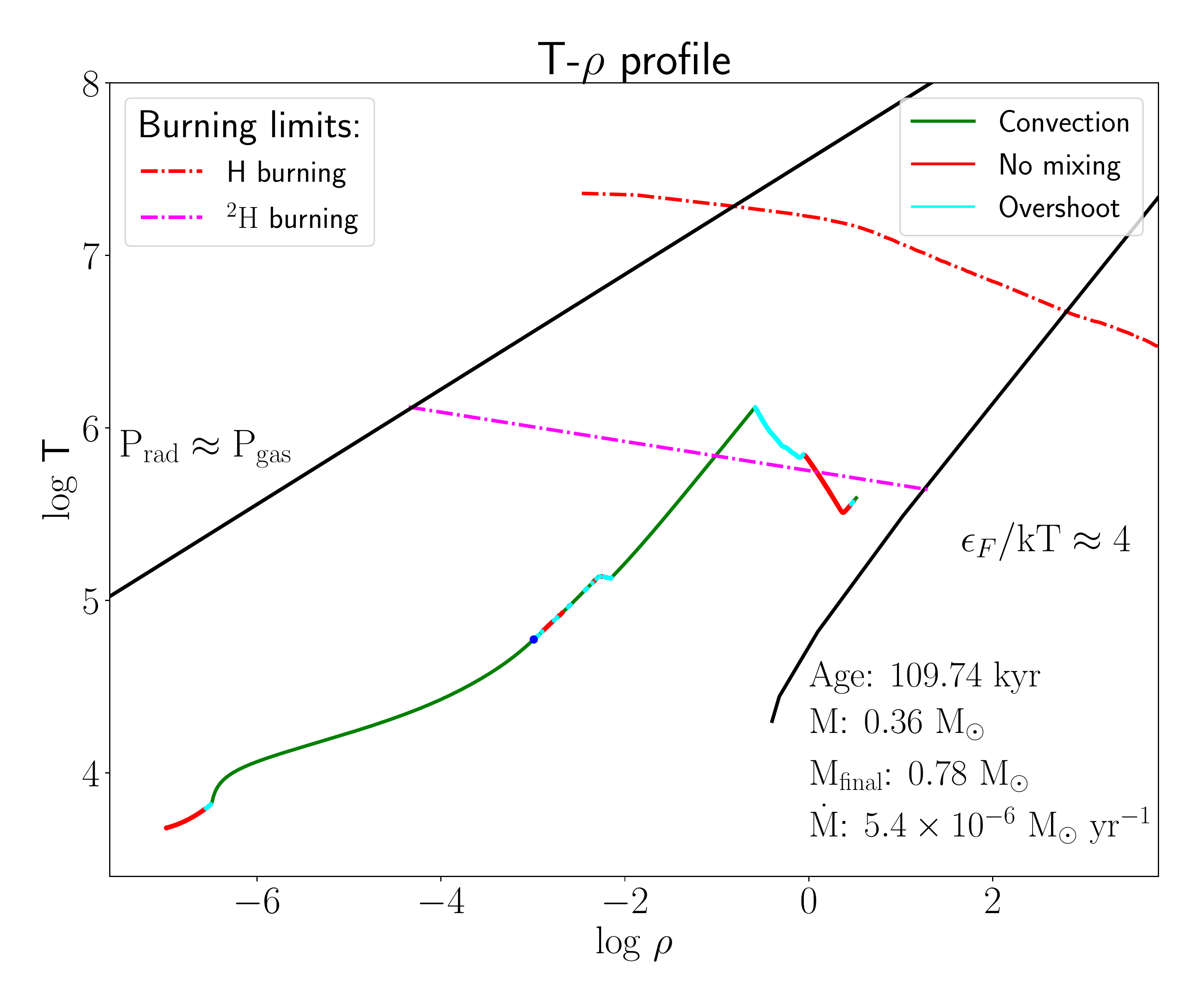}
  \caption{T-$\rho$ figure of star $\#1$ in  \Fig{thumbs}. Colours indicate how heat is transported and matter is mixed.
  Fusion limits are indicated by dash dotted lines.
  The blue point indicate the depth of the region where accretion energy is distributed at that specific point in time.
  The star was evolved using the dynamical $\alpha(\dot{M})$ model.}
  \label{fig:trho}
\end{figure}

Figure \ref{fig:sink_0025} show the evolution of the stellar variables $T_{\mathrm{eff}}$, $R$ and the luminosity from the initial accretion phase
until the ZAMS for star $\#1$; the same protostar as in \Fig{alpha-comparison}.
The initial expansion phase is clearly seen in the early stages while the effect of late accretion bursts on the evolution is evident from several bursts occurring between $100$ kyr and $1$ Myr, which halt the contraction and cause large fluctuations in the observable variables.

In Figs.~\ref{fig:alpha-comparison}, \ref{fig:sim-mass2}, and \ref{fig:sink_0025} we see instabilities near the contraction tracks towards the end of accretion.
These fluctuations in luminosity and surface temperature are driven by the deposition of accretion energy.
To investigate this in more detail, in \Fig{trho} is shown the $T-\rho$ phase-space plot for star $\# 1$.
The blue dot indicates the extent of the region where external heat from the luminosity at the accretion shock is
deposited. While it appears like a large region in phase-space, it only corresponds to 0.16\% of the mass and 7.8\% of the radius. 
The increased temperature in the outer layers creates an unstable radiative region, where the adiabatic coefficient decreases below $\gamma_{ad} < 4/3$
leading to a dynamical instability. This can be seen in \Fig{trho} where the slope of the curve drops near $\log \rho = -2$.
The region contracts releasing energy, which causes the luminosity and effective temperature to increase, while the star settles to a new stable state.
After the region has contracted the adiabatic coefficient rises and convection is restored.
A Similar instability is seen in figure 3 of \citet{2011ApJ...738..140H} (case mE-C) for a model featuring episodic accretion with a cold accretion shock
and high accretion rates of $\dot{M} = 10^{-4} \mathrm{M}_{\sun} \mathrm{yr}^{-1}$. As a consequence, in their models, a large amount of matter is added to the
outer layers of the star with the same entropy as the existing photosphere. The star responds by increasing the luminosity and effective temperature to radiate
the additional energy and reestablish hydrostatic equilibrium. The protostar does not expand since the accretion timescale, $t_{\mathrm{acc}} = \frac{M}{\dot{M}}$, is
shorter than the thermal timescale, $t_{\mathrm{KH}} = \frac{GM^2}{RL}$, which means the thermal equilibrium is not maintained. This is the case once the protostars reaches a certain mass for realistic accretion rates. The importance
of the ratio between $t_{\mathrm{KH}}$ and $t_{\mathrm{acc}}$ for accreting stars is also discussed by \citet{1985MNRAS.216...37P}. 
To assess how robust the instability is to how energy is deposited during hot accretion we have evolved models using smaller and larger regions for the energy deposition.
While this changes the details of where exactly the discontinuity occurs and the early time evolution in a burst, it does not change the overall behaviour.
Even though both \citet{2011ApJ...738..140H} and ourselves observe these instabilities, we caution that the results are based on one-dimensional stellar evolution codes.
A proper stability analysis is needed to confirm and understand the nature of the instability. 
\citet{2016A&A...588A..85G} studied dynamic 2D models of accreting protostars and found that hot accretion changes and sometimes even inverts the temperature gradient locally. This inhibits convection and leads to local dynamical instabilities as the ones described above, however it is not clear to what extent these effects are comparable.

The existence and strength of the instability vary significantly between different protostars, depending on the intermittency of their accretionary history, as
some hardly experience the instability while it dominates the PMS evolution for others as seen in Figs.~\ref{fig:alpha-comparison}--\ref{fig:sim-mass2}.
The period of the instabilities is coupled to the change in accretion rates and can be anywhere from tens of yr to a kyr. They are therefore not similar to the
observed pulsations in PMS stars, which are assumed to result from the $\kappa$ and $\gamma$ mechanisms or convective barriers
\citep[see e.g.][and references therein]{zwintz_2015}. 
As discussed below, periodically changing accretion rates in binary systems can lead to
periodic changes in the total luminosity by more than a factor of 10 on a relatively short time-scale (see \Fig{binary}).

\subsection{Comparison with Collinder 69 and the Orion Nebular Cluster}

Collinder 69 is a young open cluster located in the Orion constellation and part of the or $\lambda$-Orionis star forming region.
It has been speculated \citep{1996A&A...309..892C} that a recent supernova, less than 1 Myr ago, has exploded in Collinder 69, quenching star formation in the cluster
and creating a molecular shell at around 15 pc from the center of Collinder 69 \citep{2000A&A...357.1001L}, where ongoing star formation still continues today in cold
clouds such as B30 and B35 \citep{2015AJ....150..100K,2016ApJS..222....7L,2017A&A...597A..17H}. The young stellar objects (YSOs) in Collinder 69 includes class II and
class III stars, and stars with debris disks. The stellar population in Collinder 69 is therefore more evolved than the stellar population in our model, which
also contains new-born protostars. We have compensated for this by removing all embedded objects from our sample, as discussed below.
It is estimated to be approximately 5 Myr old \citet{2011A&A...536A..63B}, though this estimate is based on classical isochrones, which could lead
to significant errors, and \citet{2013MNRAS.434..806B} find it to be up to 10 Myr old.
Nonetheless, for simplicity and since this is the first paper to explore a self-consistent model of stellar evolution in an evolving
molecular cloud, we will consider this age to be the most likely for the cluster and compare our models at 5 Myr with the inferred
locations in the HR diagram of the members of Collinder 69 from \citet{2011A&A...536A..63B}.

The bolometric luminosities plotted in the HR diagrams are the total luminosities, which are the sum of the stellar luminosities and the outward
component of the accretion luminosities $L_{\mathrm{tot}} = L_* + L_{\mathrm{acc}}^{\mathrm{out}}$.
In the observational censuses all the embedded protostars will be absent,
and we therefore have to make an estimate for when a star is a Class 0 protostar. In principle one should do a careful evaluation of
the angle-averaged dust extinction for each star, or even compute HR diagrams from many different viewing angles of the cluster, and average
the diagrams themselves. In this paper, we do not try to make a detailed comparison between the modelled and observed clusters, and we have therefore
decided to use a simpler approach based on the correlation between envelope-stellar mass ratio and protostellar class in \citet{2016A&A...587A..59F}.
We define the envelope mass as the total gas mass inside a 10,000 AU distance from each star and compare it with the stellar mass. If the envelope
mass is larger than the stellar mass the star is assumed to embedded. Applying this selection criteria to the last snapshot in the model, we find that
182 stars are embedded at 5 Myr, leaving 413 stars from the evolved population together with 231 stars from the young population to be included
in the HR diagrams.

\begin{figure*}
\centering
   \includegraphics[width=17cm]{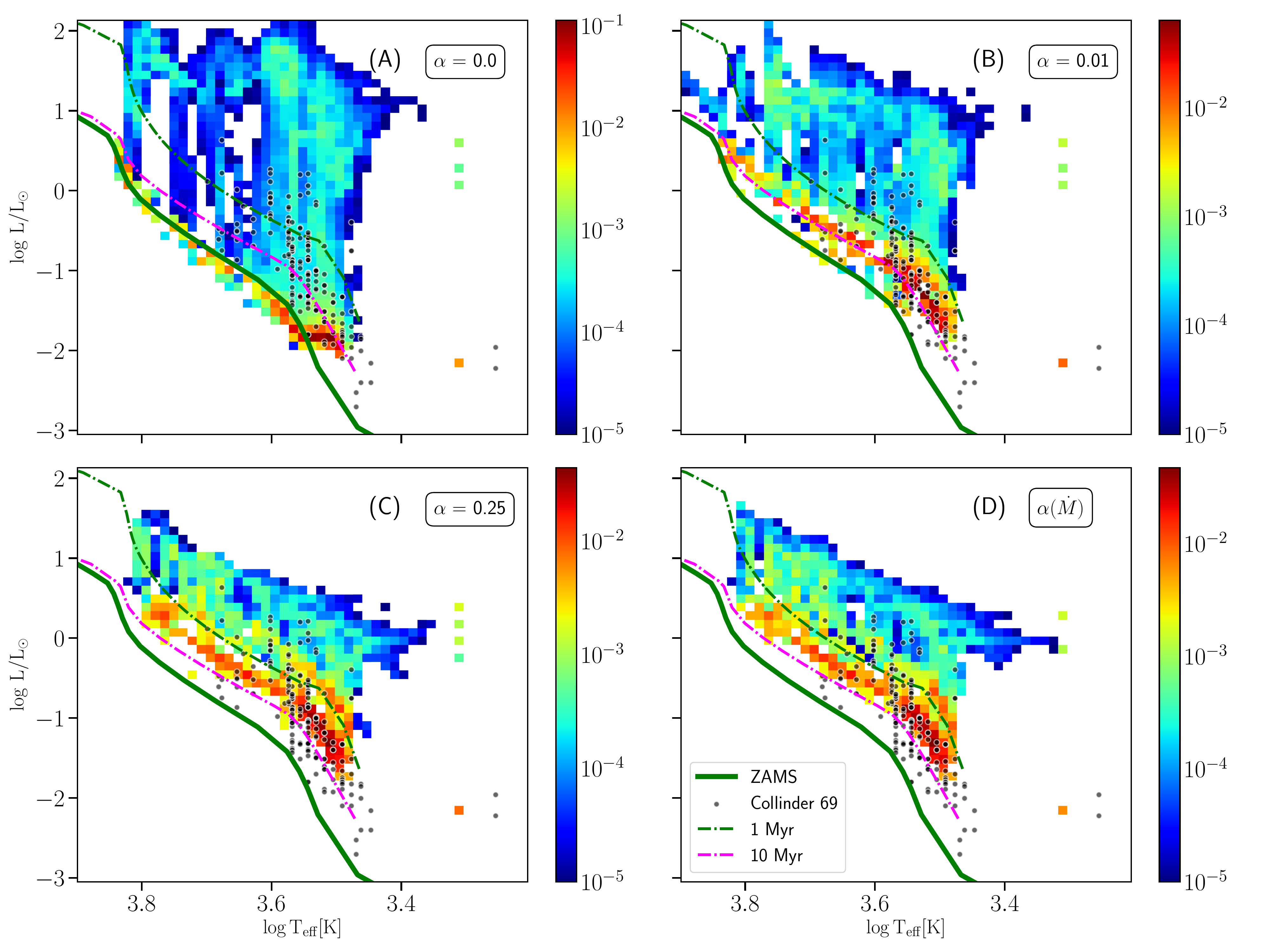}
   \caption{2D histogram of the stellar positions in the HR diagram for the thermal efficiencies
   $\alpha = 0, 0.01, 0.25, \alpha(\dot{M})$ in the synthetic cluster at a cluster age of 5 Myr.
   The density includes each position of the stellar models in the time interval $t_{\mathrm{cluster}} \in [4.9,5.1]$ Myr to
   account for the possible variations in the 200 kyr interval due to episodic accretion events. The histogram is properly weighted according
   to the time interval each star spends in the different states.
   Only stellar models with a final mass in the range $M \in [0.08, 1.5] \mathrm{M}_{\sun}$ are included. Note that the histogram
   density is log-scaled. We caution that at first sight it is easy to overestimate e.g.~the number of stars in the high-luminosity state.
   We have chosen this scaling to show the full extent of the distribution.
  Black points are confirmed
  members of Collinder 69 \citep{2011A&A...536A..63B}.
  1 Myr and 10 Myr isochrones are non-accreting pre-main sequence models evolved in \mesa{} with the same composition.}
   \label{fig:bayo_allAlpha}
\end{figure*}

\Fig{bayo_allAlpha} show an estimate of the HR diagram for PMS stars with masses in the range $M \in [0.08; 1.5] \mathrm{M}_{\sun}$ in the synthetic cluster
at an absolute age of 5 Myr. We sample the stellar distribution in a time interval of 200 kyr around 5 Myr, and construct a 2D weighted
histogram contour to show the extent of the synthetic cluster in the HR diagram. We use a 200 kyr interval instead of a snapshot, to properly
capture, in a statistical sense, any episodic accretion events that may change the luminosity significantly.
The weights corresponds to the time the star spend at that specific location in the HR diagram, and the histogram is therefore a fair
comparison with observations.
The upper mass limit is because we use a mass range similar to the estimated mass range of the confirmed members of Collinder 69.
We do not include brown dwarfs in the synthetic cluster even though they are included in the Collinder 69 data as the \mesa{} models
were not setup to handle the evolution of these, and the initial mass function in the simulation may be incomplete at these masses.
The figure includes $5$ Myr of star formation stitched together from the $2.55$ Myr molecular cloud simulation as described in section \ref{sec:model}.

\Fig{ONC_allAlpha} is similar to \Fig{bayo_allAlpha} but features members of the Orion Nebular Cluster (ONC) and the
mass is no longer limited to $1.5 \mathrm{M}_{\sun}$. For the ONC members there is a slight offset towards higher luminosities
between the high density region of the synthetic cluster and the high density region of ONC members. This could be the
result of several distinct stellar populations including a recent star formation burst \citep{2017arXiv170509496B}.
This is not entirely reproduced in the synthetic cluster simulations, due to the limited simulation time span and possibly
also the extent of the simulated molecular cloud patch.
The IMF of the ONC also show a peaked distribution somewhat different from the Chabrier IMF and the IMF in our model
\citep{2014MNRAS.444.1957D}. Younger stars created within the last Myr would appear with larger luminosities in
the synthetic cluster and could explain the discrepancy between the populations.

Regarding Collinder 69 we note that the estimated age of 5 Myr is based on the best fit to the observed data amongst the cluster members with low $T_{\mathrm{eff}}$ \citep{2011A&A...536A..63B}.
The choice to estimate the age based on the cooler cluster members is due to a smaller luminosity spread in this part of the HR diagram.
In the cooler part of the HR diagram 95\% of the cluster members lie above a 20 Myr isochrone while a larger age spread exist for the warmer members.
The mass range for the members of Collinder 69 is inferred from classical isochrones assuming a cluster age of 5 Myr.
\citet{2012A&A...547A..80B} found the mass estimates to be robust under variations in the cluster age.
We do not believe this conclusion would change significantly in a framework of accreting PMS models
due to similarities between the Hayashi tracks of accreting models and non-accreting models of similar mass.
The tracks of low- and high-mass stars do coincide at early stages for accreting models but at this stage the stars are expected to
be embedded and should thus not distort the mass estimates.

\subsection{Evolutionary stages in the HR diagram}
In a realistic model of a molecular cloud complex dominated by turbulent fragmentation massive stars are not formed immediately from
as single large core, but instead is fed through the filaments and take a significant amount of time
to accrete, reaching several Myr for the most massive stars in the simulation \citep{2014ApJ...797...32P}. This changes our picture of
stellar evolution. To illustrate it, in \Fig{evolutionary-tracks} are examples of low-, intermediate- and high-mass stellar tracks using
the variable $\alpha(\dot{M})$-model for the thermal efficiency together with arrows indicating different phases of this more dynamic picture of
early stellar evolution.

Initially, for the first $\sim$100 kyr all stars evolve in a similar manner, starting from a universal second Larson core (A).
Following almost identical tracks in the HR diagram they rapidly expand and increase their effective temperature (B).
Towards the end of the main accretion period during the deeply embedded stage, low-mass stars
start to contract, descending along the Hayashi track (C). If the star instead keeps accreting, the increase in temperature continues and
when the core becomes radiative, and deuterium burning continues in a shell, it starts to expand rapidly.
Eventually, the outer layers become radiative and the high-mass star begin to follow a Henyey track at a slightly inclined luminosity,
due to the ongoing mass accretion.
The $8 \mathrm{M}_{\sun}$ star shown in \Fig{evolutionary-tracks} reaches the Henyey track (D) at an age of $\sim$800 kyr and a mass of $\sim3.1 \mathrm{M}_{\sun}$,
and stays on it during $\sim$300 kyr. Notice that the evolution of massive stars as they join the Henyey track is almost independent of the final mass
of the star, and more dependent on how much mass they are able to accrete before the shell-burning of deuterium commence. In this case, once the
mass reaches $\sim 4 \mathrm{M}_{\sun}$ the star ignites the p-p chain in the core and soon after joins the ZAMS.
At this stage the star is massive enough that the thermal efficiency of the accretion has insignificant effect on the stellar structure and evolution.
Finally, the high-mass star in \Fig{evolutionary-tracks} during a period of $\sim$500 kyr moves on the ZAMS, almost doubling the mass and reaching
a final mass of 8 $\mathrm{M}_{\sun}$ (E).

\begin{figure*}
\centering
   \includegraphics[width=17cm]{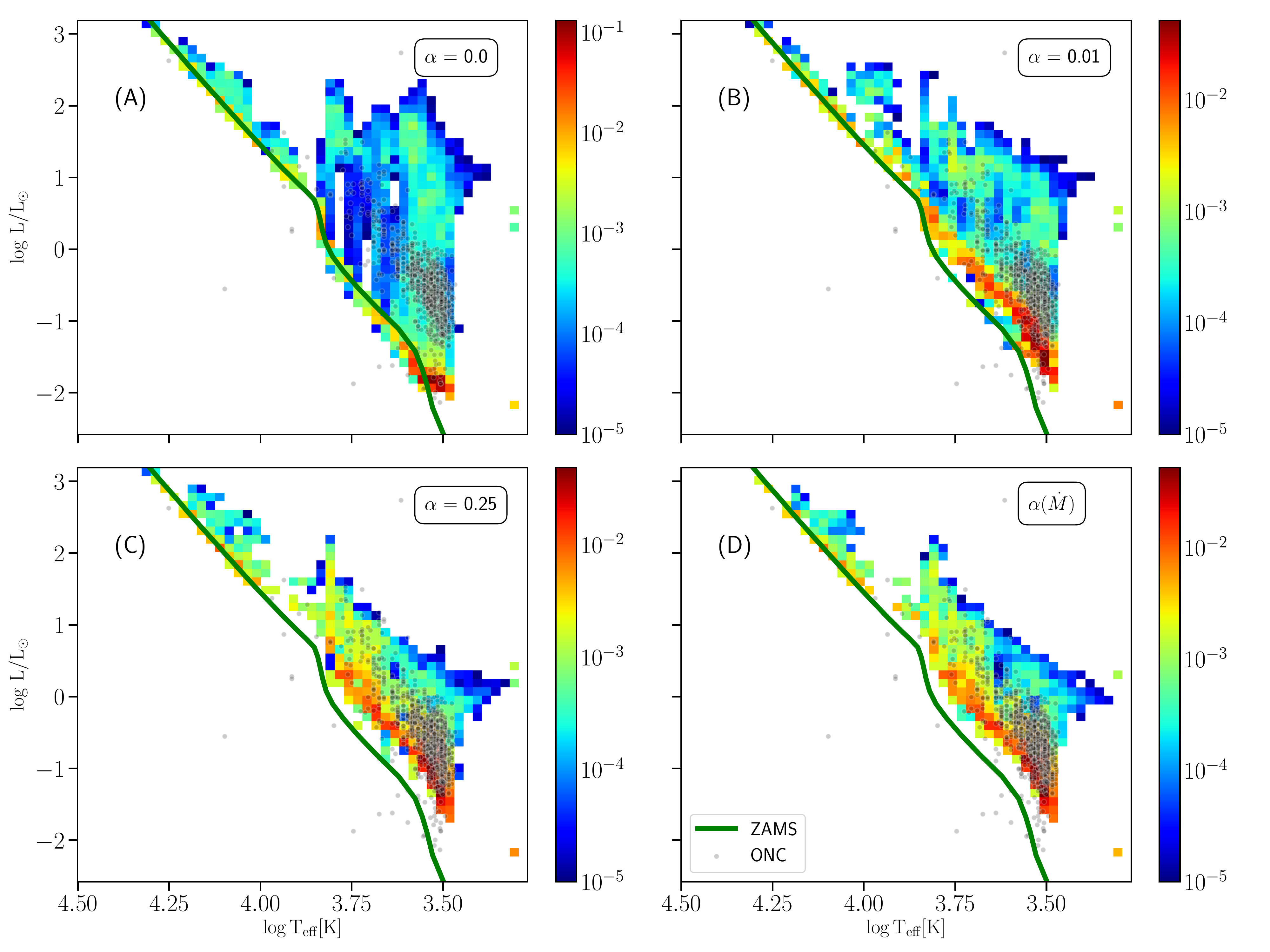}
   \caption{2D histogram of the stellar positions in the HR diagram for the thermal efficiencies $\alpha = 0, 0.01, 0.25, \alpha(\dot{M})$ in
   the synthetic cluster at a cluster age of 5 Myr.
   The density includes each position of the stellar models in the time interval $t_{\mathrm{cluster}} \in [4.9,5.1]$ Myr to
   account for the possible variations in the 200 kyr interval due to episodic accretion events. The positions are time-weighted. Note that the histogram
   density is log-scaled. We caution that at first sight it is easy to overestimate e.g.~the number of stars in the high-luminosity state.
   We have chosen this scaling to show the full extent of the distribution.   Black points are confirmed members of the Orion Nebular Cluster (ONC) \citep{2010ApJ...722.1092D}. Density is given in counts per bin.}
   \label{fig:ONC_allAlpha}
\end{figure*}

It is an important conceptual change compared to the classic picture of
non-accreting stellar evolution segmented in to low-mass Hayashi tracks and high-mass Henyey tracks. While some details may change when
the effects of e.g.~rotation, magnetic fields, outflows, and radiation are eventually included in the models, the evolution is fundamentally a
consequence of the time-scales necessary for accreting the mass, and therefore robust. The corresponding upper envelope can be appreciated
in \Fig{ONC_allAlpha}. While our tracks superficially are similar to those of \citet{2013A&A...557A.112H,2016A&A...585A..65H}, the
difference is the time-scales, since they consider rapid core-accretion, which is not supported by our global models of star forming clouds.

Exactly at which mass the high-mass stars start to move along the Henyey track depends on the early accretion rate, which is regulated by
the rate of infalling material from larger scales. In the case of a free-fall isothermal envelope this is bounded by $\sim c_s^3 / G$, where
the exact numerical pre-factor is between 1 and 50
\citep{1969MNRAS.144..425P,1969MNRAS.145..271L,1977ApJ...214..488S,1977ApJ...218..834H,1993ApJ...416..303F}, and therefore
depends on the temperature of the envelope and surrounding molecular cloud core.

The above description of the evolution applies to hot accretion models. As already discussed, cold accretion models behave very differently.
Because of the low initial temperature, the stars rapidly contract, heat up, and ignite hydrogen. Even low-mass stars reach the ZAMS as early as
after 100 kyr, and then start to evolve along the ZAMS while accreting. The late evolution, in hot accretion models, for high-mass stars along the
ZAMS is basically for the same reasons: at the point where the star reaches the ZAMS the total luminosity is completely dominated by radiation
from the surface, while the accretion luminosity is sub-dominant, and the evolution proceeds as if it was cold accretion.

\begin{figure*}
\centering
   \includegraphics[width=17cm]{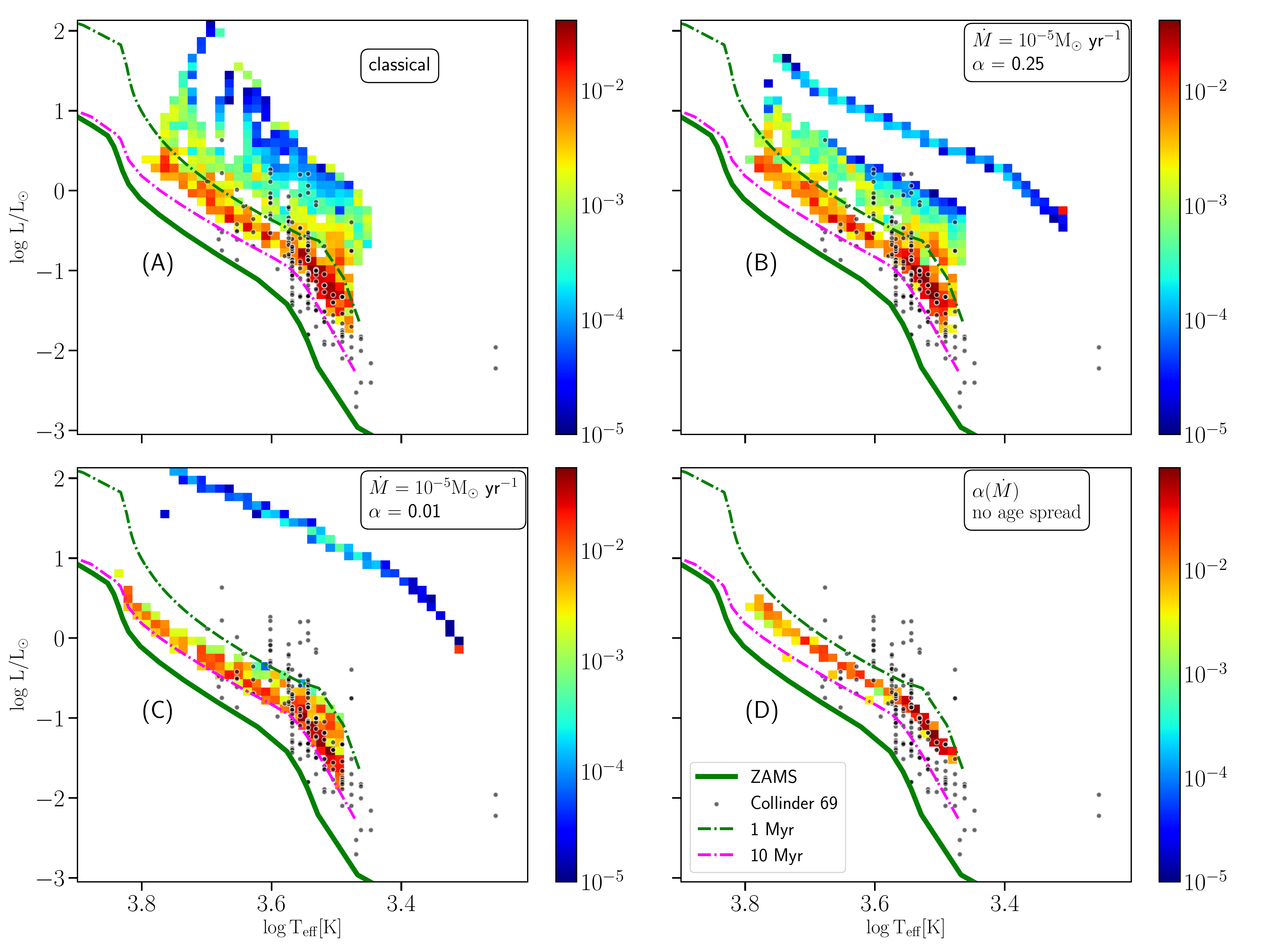}
   \caption{Similar to \Fig{bayo_allAlpha} but with different models.
   Panel A: classical PMS models evolved in \mesa{} but with the same formation history as the synthetic cluster models.
   Panel B: constant accretion rate $\dot{M} = 10^{-5}$ M$_{\sun}$ yr$^{-1}$ and $\alpha = 0.25$, again with the same formation history as the synthetic cluster.
   Panel C: constant accretion rate $\dot{M} = 10^{-5}$ M$_{\sun}$ yr$^{-1}$ and $\alpha = 0.01$, again with the same formation history as the synthetic cluster.
    Panel D: the same as panel D in \Fig{bayo_allAlpha} but without the age spread of the formation history from the synthetic cluster, meaning that all stars have evolved for 5 Myr and stopped accreting. The high luminosity tracks in panel B and C are models which are still actively accreting.}
   \label{fig:hist_alternative}
\end{figure*}

\begin{figure}
  \resizebox{\hsize}{!}{\includegraphics{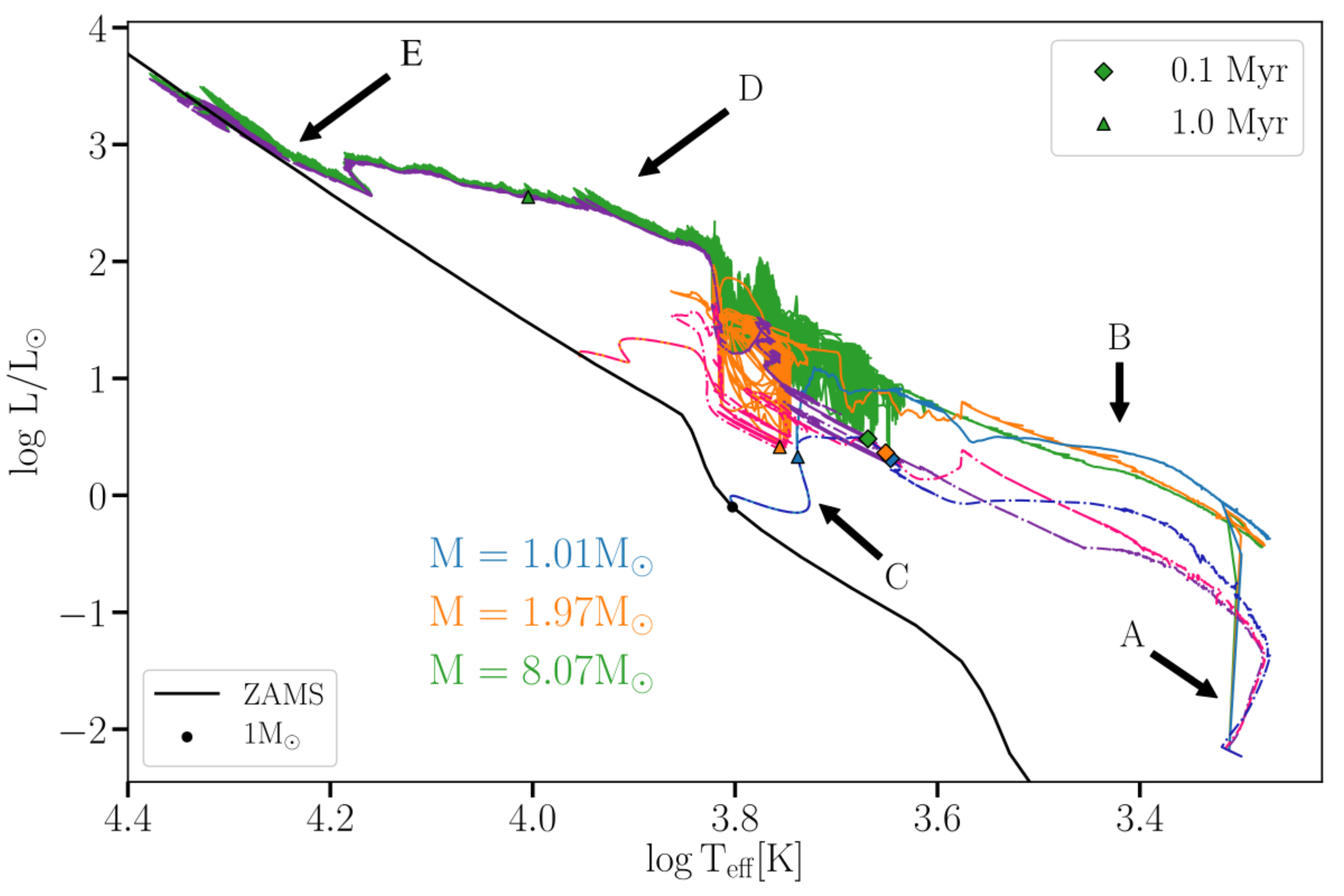}}
  \caption{HR diagram with evolutionary tracks for stars of $1$, $2$, and $8 \mathrm{M}_{\sun}$ with $\alpha(\dot{M})$. The dash dotted line
  (dark blue, pink, and purple) show the protostellar luminosity $L_*$ while the full line (blue, orange, and green) shows the total
  luminosity $L_{\mathrm{total}}$ = $L_*$ + $L_{\mathrm{acc}}$. (A) is the initial expansion phase as the accretion begins from the universal initial condition.
  (B) marks the main accretion phase, which is independent of the final mass of the star.
  (C) marks the Hayashi contraction track for low-mass stars.
  (D) marks the Henyey tracks for intermediate- and high-mass stars with continued accretion during hydrogen fusion.
  (E) marks the final accretion phase for high-mass stars along the ZAMS.}
  \label{fig:evolutionary-tracks}
\end{figure}

\subsection{Accretion burst induced by gravitational interactions in a molecular cloud environment}\label{sec:binary}
Interactions between stars on small scales and between stars and the collective gas and stellar potential on larger scales
can affect the accretion history and thereby the stellar evolution.

On the larger molecular cloud scale, as protostars move around in the collective gravitational potential, they may enter or exit high density
regions and filaments, which can cause the time evolution in the accretion rates to deviate drastically from models of isolated cores.
In Figs.~\ref{fig:cluster} and \ref{fig:thumbs} we see that some stars cluster together in dense regions while others evolve quietly in isolation
\citep[see also][]{kuffmeier}.
The accretion profiles show large variations with a general trend where unsteady accretion rates are associated with stars in dense regions
harboured in multiple systems, while isolated stars have smoother and monotonically decreasing accretion profiles.

\begin{figure}
\resizebox{\hsize}{!}{\includegraphics{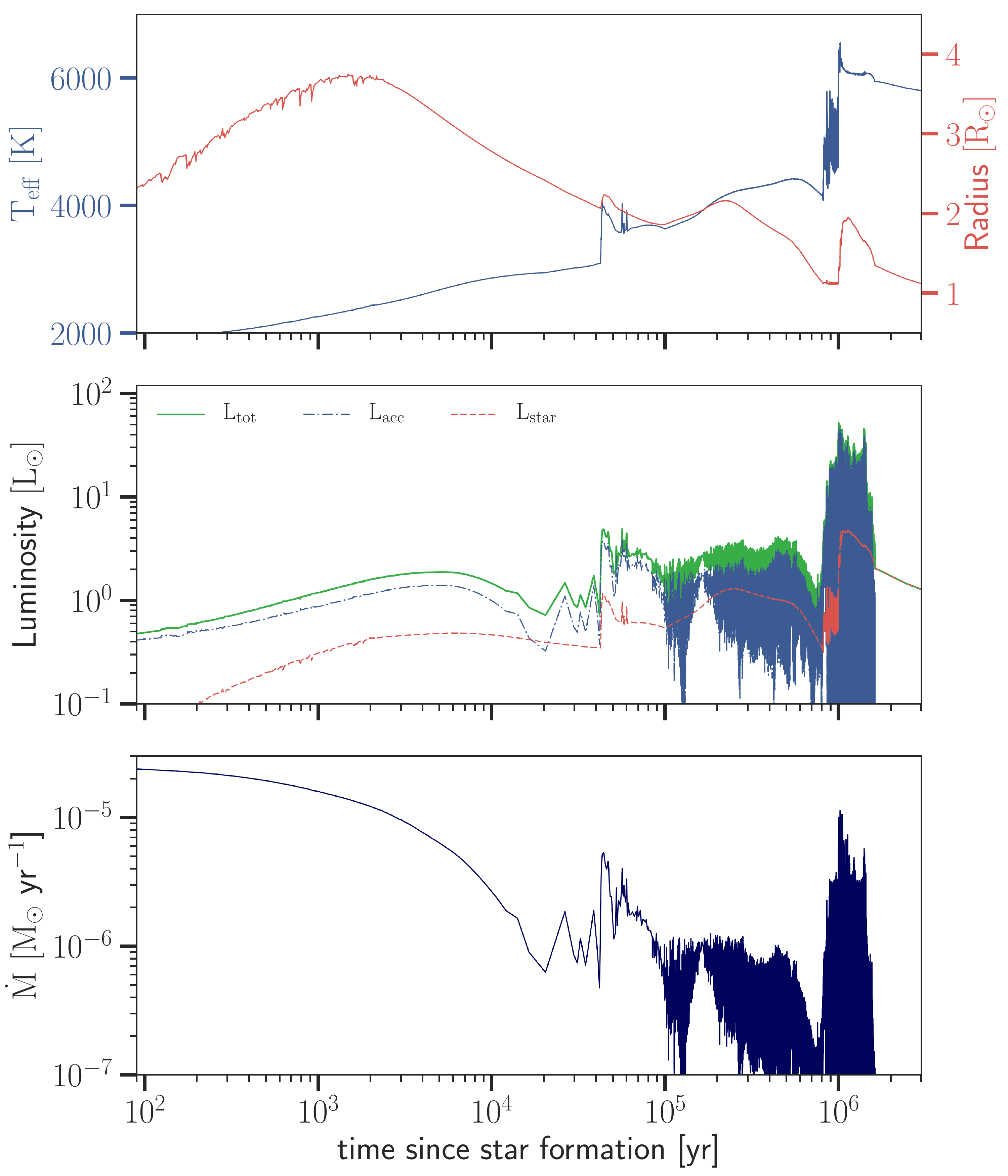}}
   \caption{Example of a protostar with periodic accretion rates induced by binary interactions.
   Top panels shows the evolution of $T_{\mathrm{eff}}$ and $R$.
   Middle panel show the total luminosity as well as the two component $L_{\mathrm{acc}}$ and $L_{\mathrm{star}}$.
   Lower panel shows the accretion rate.
   \Fig{binary_closeup} show a smaller timespan where
   the periodicity is clearly visible.
   \label{fig:binary}}
\end{figure}

\begin{figure}
\resizebox{\hsize}{!}{\includegraphics{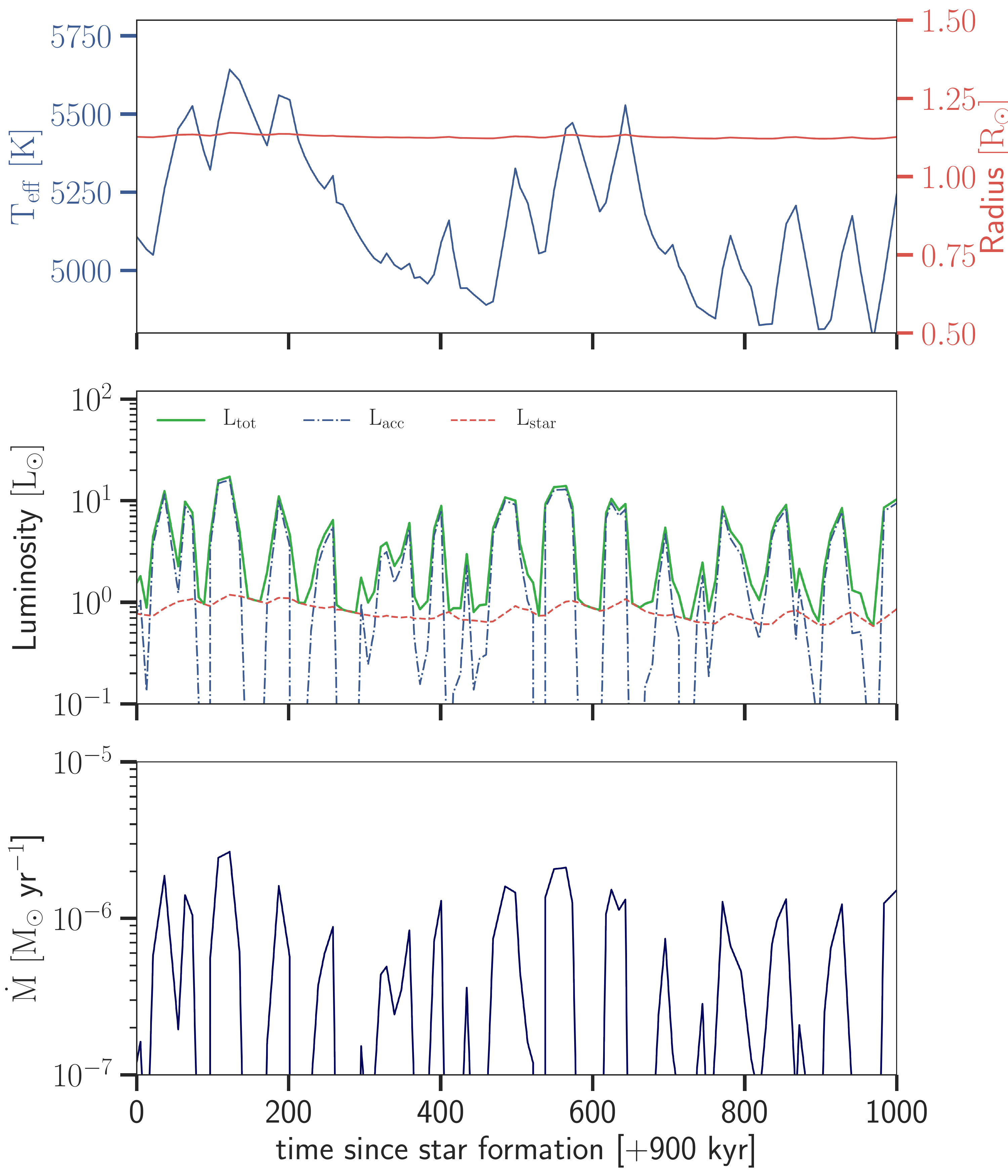}}
   \caption{Same panel as \Fig{binary} but with a closeup over a 1000 yr period highlighting the periodicity in the accretion rate.
   \label{fig:binary_closeup}}
\end{figure}

Approximately half of all young solar mass stars are found in binary or multiple systems \citep{2016ApJ...818...73T}.
This can induce periodically varying accretion rates \citep{2014ApJ...797...32P} and affect the protostellar and PMS evolution.
In Figs~\ref{fig:binary} and \ref{fig:binary_closeup} we show an example of a protostar with a binary induced periodicity in the accretion rate.
At an age of 900 kyr the shown star has a mass of 0.5 $\mathrm{M}_{\sun}$, while its primary companion is a 1.3 $\mathrm{M}_{\sun}$ star. The system has a
semi-major axis of $a=19.5$ AU and an eccentricity of $e=0.76$. The period is 63.5 yr.
This periodicity gives large fluctuations in the effective temperature and halts the contraction of the protostar, thus delaying the evolution
towards the ZAMS. The periodic accretion bursts also lead to luminosity fluctuations of more than one order of magnitude in sync with the orbit time.
This is because the secondary is on an elliptic orbit, where at aphelion it has the smallest velocity difference with the envelope, and therefore the
largest accretion rate.

As a result it is not readily evident to an observer at what evolutionary stage the star is in solely based on the accretion rate, and a
binary system can have luminosity variations on the same scale as FU Orionis-like bursts with time-scales of tens to hundreds of years
and periodic in nature. Such bursting binary systems could occur both in the protostellar and the PMS evolutionary phases. Recently, several
deeply embedded binary protostars have been found with a relatively small seperation and clear chemical fingerprints of recent episodic accretion
events \citep{2017arXiv170310225F}.

\section{Discussion}
\subsection{Accreting protostellar models as a solution to the observed luminosity spread}
One motivation behind the development of accreting protostellar models have been to
address the observed luminosity spread in young stellar clusters as proposed by \citet{2009ApJ...702L..27B}.
Figs~\ref{fig:bayo_allAlpha} and \ref{fig:ONC_allAlpha} show good agreement between the luminosity spread
in the observed HR diagrams of two young clusters and the luminosity spread in the synthetic cluster
if one allows some variation in $\alpha$ values in a coeval population.
The luminosity spread in the synthetic cluster is not solely a consequence of the episodic nature of the accretion profiles
or the implementation of the thermal efficiency of the accretion shock, but is the result of a number of different factors.

The variations in the thermal efficiency of the accretion shock produce large variations in the protostellar evolutionary tracks,
but the effect is mostly limited to the protostellar phase ($t_{\mathrm{star}} < 1$ Myr), where variations in
accretion rates in general are most pronounced causing large fluctuations in the luminosity and radii of the individual models.
Once the protostars enter the PMS characterised by low accretion rates with occasional accretion bursts,
the variations are much less pronounced.
For the warm models with $\alpha > 0.1$ the tracks for different thermal efficiencies converge towards each other and the memory
of the initial accretion history and the differences in thermal efficiency is essentially erased.
From \Fig{alpha-comparison} we see that these models reach the $1$ Myr point at almost identical positions in the HR diagram
and follow similar tracks towards the ZAMS.
The PMS tracks of the warm models are very similar to the classical Hayashi tracks stars with the same masses.
In the purely cold models with $\alpha = 0.0$ stars follow very different tracks and have much lower luminosities,
however as discussed in section \ref{sec:HRtracks_alpha} we do not believe the evolution of the purely cold models, with $\alpha = 0.0$ throughout the entire accretion phase, are representative of the majority of real protostars.
In \Fig{alpha-comparison} we see that the model with $\alpha = 0.01$ has a much shorter contraction phase than the warmer models.
Models with a thermal efficiency in the range $0.0 < \alpha < 0.10$ can produce a substantial spread in luminosity as the length of the contraction tracks and the temporal evolution differ. Thermal efficiencies in this range are necessary to explain the faintest members in Collinder 69 as seen in panels A and B in \Fig{bayo_allAlpha}.
The shorter contraction phase of the stellar models with low thermal efficiency means that these would appear older
in the old paradigm of non-accreting isochrones, where the relatively low luminosity of these models
would not be reached until late in the PMS phase.
Comparing the non-accretion $10$ Myr isochrone from \citet{2015A&A...577A..42B} with the lower boundary of the accreting models with non-zero $\alpha$ in
\Fig{bayo_allAlpha} the accreting models would appear much older if the non-accreting isochrones was used to estimate the age.

Allowing for significant variations in thermal efficiencies between coeval stellar populations can produce an additional luminosity spread. In figure \ref{fig:bayo_allAlpha} a combination of panels B and D in would almost reproduce the entire Collinder 69 luminosity spread.
However this combination requires that some stars have a low thermal efficiency throughout the entire evolution
 including the initial main accretion phase where accretion rates usually exceed $\dot{M} = 10^{-5}$ M$_{\sun}$ yr$^{-1}$. It is uncertain if this is physically feasable.
The purpose of the dynamical $\alpha$ model was to replicate a combination of several $\alpha$ values, however our model does not suffice in this regard as all the protostars undergo hot accretion at some stage.
It should be noted that the maximum thermal efficiency presented here, $\epsilon = 0.5$ \& $\alpha = 0.5$,
is a rather conservative upper limit for the thermal efficiency compared with the theoretical maximum
efficiency of $75\%$ from \citet{1980ApJ...241..637S}.
Additionally, we do consider it likely that some of the observed cluster members in figs~\ref{fig:bayo_allAlpha} and \ref{fig:ONC_allAlpha}
are unresolved binaries, which will have a higher apparent luminosity than the individual stars.

The introduction of realistic accretion histories create large variations in the evolutionary tracks
in the early stages of accretion, but the effects are again limited after $t_{\mathrm{star}} > 1$ Myr. It should
be noted though, that due to the limited resolution in our model (50 AU), we are only capturing
the infall rate of material from the envelope to the disk, and using it as a proxy for the the accretion rate
of material from the disk to the star. Given sufficient numerical resolution and relevant microphysics in the model to resolve and correctly evolve
the disks around the stars, a higher intermittency in time and large variability in amplitude of the accretion rates are to be expected \citep{kuffmeier}.

To construct a model that captures the probable differences in the accretion flow as a function of the accretion rate, we have introduced
the dynamic thermal efficiency $\alpha(\dot{M})$, which further enhances the significance of the accretion histories.
In \Fig{sim-mass1} and \Fig{sim-mass2} we illustrate differences in the evolution tracks of stars with the dynamical thermal efficiency and varying accretion rates.
We see that there exists minor temporal variations between the models as they reach the PMS contraction tracks, with a $\sim 0.75$ dex spread of the $1$ Myr markers.
These variations go some way towards explaining the observed luminosity spread in young clusters, but overall the PMS evolution
of the models with the same thermal efficiency is not that different and the effect of varying accretion histories is not sufficient to account for the whole luminosity spread.

From the global molecular cloud simulations we obtain a realistic age spread of protostars in a coeval population,
something which has not previously been included in studies of accreting protostellar models in relation to the stellar structure and reproducing
the luminosity spread in the HR diagram.
Including the absolute formation time of each protostar naturally increases the spread in luminosities as some stars are much younger
than the age of the cluster itself. This effect is important for replicating the observed luminosity spread.

In \Fig{hist_alternative} panels A,B, and C we show classical PMS models and constant accretion models evolved in \mesa{} following the same
birth times and stellar masses as the protostars of the global molecular cloud models.
Panel D shows the dynamical $\alpha$ model with the realistic accretion rates but excluding the intrinsic age spread from the cluster simulation. Removing the age spread all the stars are presented at an age fo $5 \pm 0.1$ Myr. At this stage accretion is ceased and, as discussed above, the luminosity variations caused by the different accretion profiles is significantly diminished.
This illustrates the importance of the natural age spread of the coeval stellar populations. When the natural age spread is considered some stars are actively accreting at different rates while others are past the accretion phase and this effect naturally contributes to a significant luminosity spread.
Both the classical and the constant accretion models produce a decent luminosity spread when the stellar age spread is included.
Comparing figs.~\ref{fig:bayo_allAlpha} and \ref{fig:hist_alternative} we see that ultimately the realistic accretion rates does increase the luminosity spread and fit the observed data better. Specifically comparing the panel B in \Fig{bayo_allAlpha} with panel C in \Fig{hist_alternative} the realistic accretion rates produce more faint objects whereas that models with similar $\alpha$ but constant accretion rate fail to produce objects with luminosities similar to the faintest objects in Collinder 69. The variable accretion rates also produce more high luminosity objects at a cloud age of $5$ Myr which is due to the longer accretion timescales of the accretion profiles obtained from the molecular cloud simulation.

The global models also introduce intrinsic effects of the environment to the accretion rates.
The local density and gravitational perturbations between protostars in star forming regions provide variations in accretion histories,
which can provide fluctuations in luminosity beyond the main accretion phase.
Such variations are evident in figure \Fig{thumbs}, where the extend of the accretion profiles vary significantly between different
protostars of similar final mass.
Some exhibit large accretion burst beyond the $1$ Myr mark, while others have finished accretion in less than $\sim500$ kyr.
Binary interactions are also visible for some of the accretion profiles as discussed in section \ref{sec:binary}.
Accounting for these variations requires global simulations with high dynamical range in both space and time. In particular, secondary
stars in binary systems on elliptic orbits will preferentially accrete the gas when close to the aphelion, where they spend most of the
time and the velocity differential with respect to the gas is the smallest. This induces regular periodic and sometimes dramatic changes
in the accretion rate \citep[see also][]{2014ApJ...797...32P}. In the case of short-period orbits in a deeply embedded core where accretion
rates can reach $\sim10^{-5} \mathrm{M}_{\sun}$ yr$^{-1}$ in the actively accreting state, this ``forcing'' could induce variability in the stellar
structure with periods equal to the binary period creating an externally forced variable star as seen in figs.~\ref{fig:binary} and \ref{fig:binary_closeup}.

\citet{kuffmeier} presents zoom-simulations for 6 protostars from a similar molecular cloud simulation with an increased spatial
resolution down to 0.6 AU.
At this resolution the accretion disk and accretion bursts driven by disk instabilities are better resolved.
These accretion burst can further increase the luminosity spread of the accreting protostellar models and increase the dependence on the thermal efficiency.

In recent works, \citet{2017A&A...597A..19B} and \citet{2017arXiv170600502V} have studied accreting PMS stars
with accretion histories from two-dimensional hydrodynamical disk simulations, which feature accretion burst
driven by gravitational instabilities within the disk.
These models complement our models as they are able to resolve disk instabilities better than the
global molecular cloud simulations utilised here.
However they lack the additional effects from the local cloud environment, the ability to track the absolute cluster age
and both ideal and non-ideal magnetohydrodynamical effects, which could alter the circumstellar disk structure and the characteristics
of the accretion bursts. 
Both groups also experiment with varying thermal efficiencies as well as hybrid $\alpha$ models similar to the $\alpha(\dot{M})$ model presented here.
\citet{2017A&A...597A..19B} find a bimodal luminosity distribution in their HR diagrams with the cold models producing very faint objects and the hybrid models producing brighter objects which lie close to the classical isochrones. To reproduce the luminosity spread they suggest it is necessary to allow varying values of $\alpha$. 
These conclusions are in accordance with our findings where a combination of panel B and D in \Fig{bayo_allAlpha} would produce the best fit to Collinder 69.
\citet{2017A&A...599A..49K} recently investigated the effects and implications of varying deuterium mass fractions in
accreting models of PMS stars.
They found that variations in the deuterium mass fraction have insignificant effects on hot accreting
models, while cold or slightly warm accretion models can exhibit large evolutionary variations depending on the deuterium abundance.
To which degree larger variations in abundances within a cluster is feasible depends on the amount of mixing and pollution
by stellar winds and exploding stars in the environment \citep{2013ApJ...769L...8V}.

In summary, the HR diagram of the synthetic cluster populations are in good agreement with the HR diagrams of
observed young cluster as is seen in Figs.~\ref{fig:bayo_allAlpha} and \ref{fig:ONC_allAlpha}, especially if we allow some protostars to accrete with a
low thermal efficiency throughout the entire formation. This agreement would most probably only improve, if we were able
to capture the additionally variability in the accretion rate induced by circumstellar disks in the model.

Developing isochrones to improve the age estimation of young clusters and PMS stars is beyond the scope of this paper, but we suggest the
following method as a viable way to improve cluster age determinations in future work.
First, a stellar population from the synthetic cluster simulation is sampled $N$ times\footnote{where $N$ is chosen to be sufficiently large} for a specific cluster age.
The stellar population should be similar to that of the observational data, e.g the mass should be in the correct mass bracket and number of stars should be the same.
Subsequently, each sample is compared to the observed positions in the HR diagram using a Kolmogorov-Smirnov (KS) test
to determine to what confidence level we can reject the null-hypothesis that the underlying distributions match.
Choosing a reasonable confidence level, the number of rejections is calculated and used to estimate the whether the synthetic cluster age
is a good match for the observed cluster population.
This procedure is repeated for a number of synthetic cluster ages and the cluster age with the least rejections is consider to be the most likely.
Such a procedure could provide a solid age estimates for young clusters, but the choice of thermal efficiency would remain an free parameter, something which
future work would need to address.
A similar statistical approach could also be developed to estimate the mass of PMS stars in young clusters, and the probability that they are in a
high-luminosity state. This could be done using the distribution of stellar models that are contributing to model pixels overlapping with the
observed star in the HR diagram.

\subsection{Constraining the thermal efficiency}
The development of realistic co-evolved models of stellar evolution is important not only in the context of star formation itself,
but also in relation to the study of protoplanetary disk structure, planet formation and the early solar system as improved models of
irradiation play a vital role in the physics and chemistry of circumstellar disks.
In \Fig{alpha-comparison} we see that large variations in thermal efficiency $\alpha$ have a profound effect on the stellar
luminosity as well as the effective radius and temperature, primarily in the early stages of star formation.
Recent observations have shown that circumstellar disks appear while the protostar is still embedded increasing the
importance of a realistic protostellar structure model for a proper account of photo chemistry, dust evolution, and the
general development of models of circumstellar disks \citep{2013A&A...560A.103M, 2017ApJ...834..178Y}.

We have compared observations of two young clusters with a synthetic cluster with varying thermal efficiencies and found that modest thermal efficiencies
are sufficient to produce comparable luminosity spreads in the synthetic cluster.
Obtaining bounds on $\alpha$ with the models and data presented here remains difficult as uncertainties in observational methods as well as
simplifications in the numerical models cause fine tuning of the efficiency to be inconclusive at this point.
Looking at panel B in \Fig{bayo_allAlpha} the models with $\alpha = 0.01$ fit the observed data with low luminosities very well. Due to accretion bursts,
higher luminosities can also be explained at this low thermal efficiency, though the density of observed stars at higher luminosity suggest that some stars must
have at least some periods of higher thermal efficiency or be younger than the modelled population to reproduce the observed luminosity spread.

The dynamical model $\alpha(\dot{M})$ gives a reasonable fit to the observations and can be used as a phenomenological
approximation until better models for the thermal efficiency are developed. It should be stressed that even though we believe
the model could be a good approximation given the general understanding we have of protostellar
accretion, it does not have a strong quantitative theoretical basis.

As noted by \citet{2017A&A...597A..19B}, constraining the thermal efficiency is a formidable problem of radiative hydrodynamics,
which neither theoretical nor observational approaches have been able to solve yet.
In a recent paper, \citet{2017ApJ...836..221M} studied the problem in the context of planet formation
using one-dimensional radiative hydrodynamics to model the shock above a hydrostatic planetary atmosphere.
Their model does not include the effects of higher dimensionality or magnetic fields, but provides a first step towards
solving the problem with numerical models.
Assuming the dynamics is comparable to the protostellar case we can use these results as a guideline for
reasonable thermal efficiencies.
They found that the thermal efficiency varies depending on accretion rates, planetary mass and shock radii, which
shows that large variations in efficiency can occur depending on the specific properties of the protostar and the local environment.
They also find that even if all the kinetic energy is effectively converted to radiation at the shock,
the accreting matter can trap a large fraction of the energy, which will then be reaccreted.
If this is the case, purely cold accretion does not seem possible during periods with high accretion rates, which is in line with the
conclusions we have made based on our cold models.
These results seems to favour a model where high accretion rates result in higher thermal efficiencies as
is the case with the dynamical model presented here.

\section{Conclusion}
In this exploratory paper we have presented a new methodology for determining ages of stellar associations by using synthetic clusters
simulations combining accretion histories from global molecular cloud simulations with
accreting protostellar structure models with varying thermal efficiency.
The molecular cloud simulation has provided us with realistic accretion rates accounting for the dynamical nature of the star forming region as well as the complete star formation history within the cloud enabling a more realistic comparison to observations of young clusters. Ongoing star formation in the molecular cloud simulation, in accordance with what is seen observationally for the Lambda-Orionis star forming region and the ONC, is important for a sufficient age spread in the stellar population, and the presence of young stars that are still undergoing periods of increased accretion.

We have experimented with a number of different thermal efficiencies including a dynamical model dependent on the instantaneous
accretion rate.
We find that protostars evolved with cold accretion throughout the entire accretion process are less likely to be compatible with observations of young stellar objects, however a small fraction protostars evolved with insignificant accretion energy are necessary to explain the faintest observations. Meanwhile a range of non-zero thermal efficiencies are in
good agreement with majority of the observed cluster members.
Hot accretion with thermal efficiencies above 0.5 does not appear to be required to explain the observed HR diagrams of young clusters.
Given the many competing factors that enters into creating a specific synthetic HR diagram, constraining the thermal efficiency further
is not currently possible.

We construct a dynamical model for the thermal efficiency depending on the accretion rate $\alpha(\dot{M})$. 
This models cannot explain the entire luminosity spread by itself, but in combination with models with low thermal efficiencies of $\alpha \sim 0.01$ the observed luminosity spread of the young cluster Collinder 69 is reproduced within the estimated $5$ Myr age.
We also find reasonable agreement between the luminosity spread of the Orion Nebular Cluster and the synthetic cluster 
though the former show signs of several star bursts over a larger volume, which is not the case for the synthetic cluster.
Extending the run time, size and mass of the global molecular cloud simulation model to be closer to that of the ONC could
possibly improve the concordance between synthetic models and the ONC.

In binary systems where the secondary is on an elliptic orbit the infall from the envelope can induce variability in the stellar
structure with periods equal to the binary period creating an externally forced variable star. The changes in total luminosity can be
of up to two orders of magnitude, which is close to what is seen for both embedded protostars and FU~Orionis like stars.

The large time-scales involved in massive star formation and the existence of a universal initial condition changes the evolution
history of massive stars, with a relatively large time-span first evolving diagonally in the HR diagram, then on a slightly inclined
Henyey-track due to the ongoing accretion, and finally moving along the ZAMS, while reaching the final mass.

In conclusion, to obtain precise ages for young stellar clusters we need to go beyond simple isochrones and consider
the combined impact of (i) realistic, time-dependent, accretion rates, (ii) warm accretion and (iii) the formation time of individual stars
This can be done using stellar evolution models obtained from simulations of star forming regions. Only then the observed luminosity
spread can be fully accounted for.
The numerical data sets used in this publication, including the stellar structure evolution and gas properties in the vicinity of the protostars, are available upon request to the authors.

\section*{Acknowledgements}
We would like to thank Remo Collet, S{\o}ren Frimann, G\"unter Houdek, Jes J{\o}rgensen, Michael Kuffmeier, Paolo Padoan,
and Mads S{\o}rensen for valuable comments and discussions while carrying out this work.
We are grateful to the anonymous referee for constructive comments that helped to improve the manuscript.
TH was supported by a Sapere Aude Starting Grant from the Danish Council for Independent Research.
Research at Centre for Star and Planet Formation is funded by the Danish National Research Foundation (DNRF97).
We acknowledge PRACE for awarding us access to the computing resource CURIE based in France at CEA for
carrying out part of the simulations.
Archival storage and computing nodes at the University of Copenhagen HPC center, funded with a research
grant (VKR023406) from Villum Fonden, were used for carrying out part of the simulations and the post-processing.

\bibliographystyle{mnras}
\bibliography{refs} 

\bsp	
\label{lastpage}
\end{document}